\documentclass[preprintnumbers,showpacs,amsmath,amssymb,floatfix,9pt,prd,onecolumn,
superscriptaddress,nofootinbib]{revtex4}
\usepackage[pdftex]{graphicx}
\usepackage{latexsym}
\usepackage{epsfig}
\usepackage{amssymb}
\usepackage[utf8]{inputenc}

\begin{document}
\title{Phantom energy supported wormhole model in $f(R,\,T)$ gravity assuming conformal motion}

\author{Piyali Bhar\footnote{Corresponding author}}
\email{piyalibhar90@gmail.com; piyalibhar@associates.iucaa.in}
\affiliation{Department of
Mathematics,Government General Degree College, Singur, Hooghly, West Bengal 712 409,
India}

\author{Pramit Rej}
\email{pramitrej@gmail.com
 } \affiliation{Department of
Mathematics, Sarat Centenary College, Dhaniakhali, Hooghly, West Bengal 712 302, India}

\author{P. K. Sahoo}
\email{pksahoo@hyderabad.bits-pilani.ac.in} \affiliation{Department of Mathematics, Birla Institute of Technology and Science-Pilani, Hyderabad Campus, Hyderabad 500078, India}

\begin{abstract}

In this article, we have discussed Morris and Thorne (MT) wormhole solutions in a modified theory of gravity that admits conformal motion. Here we explore the wormhole solutions in $f(R,\,T)$ gravity, which is a function of the Ricci scalar ($R$) and the trace of the stress-energy tensor ($T$). To study wormhole geometries, we make assumption of spherical symmetric static spacetime and the existence of conformal Killing symmetry to get more acceptable astrophysical outcomes. To do this, we choose the expression of $f(R,\,T)$ as $f(R,T)= R+2 \gamma T$. Here we employ the phantom energy EoS relating to radial pressure and density given by $p_r=\omega \rho$ with $\omega<-1$ to constrain our model. Following a discussion of wormhole geometry and behavior of shape function, the study moves on to the computation of proper radial distance, active mass function, the nature of total gravitational energy and a discussion on the violation of energy conditions. We have shown that the wormhole solutions exist for positive as well as negative values of the coupling constant $\gamma$. From our analysis we see that no wormhole solution exists for $\gamma =-4\pi,\,-\pi(3+\omega)$. All the physical parameters have been drawn by employing the values of $\gamma$ as $\gamma=-0.3,\,-0.2,\,-0.1,\,0,\,0.1$ and $0.2$, where $\gamma=0$ corresponds to general relativity (GR) case. It is found that for our proposed model, a realistic wormhole solutions satisfying all the properties can be obtained.

\end{abstract}

\maketitle

\section{Introduction}

Wormholes (WHs) are hypothetical geometrical structures that connect two different space-times or two different points within the same space-times. Wormholes are asymptotically flat tubelike structures. For their properties, they are assumed to be very useful for interstellar travel. The concept of wormholes is the most well-known research topic both in general relativity, as well as in modified theories of gravity. In the literature, these are classified into two categories: static wormholes and dynamic wormholes, depending on whether the radius of the throat of wormhole is constant or variable in nature. The throat of a wormhole is the tube-shaped smallest surface area linking the two universes between the two mouths of the WH. Despite extensive research, its nature is still not fully understood. WHs emerge from general relativity solutions (GR). The term `wormhole' was first coined by Flamm \cite{Flamm1916}. Flamm first obtained a WH-like solution in the form of Schwarzschild’s WH. It was later discovered that it would collapse extremely fast, making it impossible to traverse \cite{Fuller1962}. Following Flamm, Einstein and Rosen obtained a similar geometrical structure known as the Einstein-Rosen bridge \cite{Einstein1935}. Weyl \cite{Weyl1921} also proposed a wormhole model in connection with mass analysis of the Einstein’s field equations. At the Planck scale, Wheeler discovered Reissner-Nordstr\"om (RN) or Kerr wormholes in the form of quantum foam geometries \cite{Wheeler1962}. Hawking \cite{Hawking1988} transformed these wormholes into Euclidean wormholes, which made it impossible to travel from one side to the other. Later in one of the pioneering work, Thorne with his student Morris \cite{Morris1988} first explored the idea of traversable wormholes as a medium for teaching GR. They performed a meticulous study by considering static and spherically symmetric wormholes. They demonstrated that the exotic matter within them contains negative energy, thereby violating the null energy condition (NEC). They proposed a class of wormhole solutions to the Einstein field equation that indicated how humans could travel through the tunnel of the wormholes as long as there was no horizon across the tunnel. It is one of the most active areas of research in this field in recent years and researchers have been trying to build accurate wormhole models with the prospect of minimizing or perhaps fully curing this violation. The principle of GR, which predicts that the geometry of space-time is distortable when the matter is present, was utilized to discover the possibility of time travel. Exact solutions connecting two separate verges of asymptotically flat space-time, as well as two unique anti-de Sitter regions, were derived in refs. \cite{Morris1988,Visser1995,Lemos2003}. The concept of wormholes is still theoretical, though further researches are being done toward achieving a WH model \cite{Jordan1959,Nandi2006,Harko2009,Bambi2013,Kuhfittig2013,Rahaman2013,Li2014,Tsukamoto2016,Zhou2016,Bhar2016a,Tsukamoto2017}. These wormholes can also be transformed into time machines for backward time travel \cite{Morris1988a,Frolov1990}.\par

The validation of energy conditions in the context of wormholes is an important topic that has been addressed in the literature, for example in dynamic and thin shell wormholes, by presenting innovative methodologies \cite{Kar1995,Arellano2006,Cataldo2011,Mehdizadeh2015}. Aside from that, various researchers have attempted to find wormhole solutions utilizing modified theories of gravity as a foundation. These wormhole solutions are developed in Brans-Dicke theory, Born-Infeld theory, Lovelock gravity, Kaluza-Klein gravity, Einstein-Gauss-Bonnet theory, Einstein-Cartan theory, scalar-tensor theory, etc \cite{Agnese1995,Nandi1997,Dzhunushaliev1998,Bronnikov2004,Richarte2009,deLeon2009,Lobo2010,Eiroa2012,Zangeneh2015,Shaikh2016,
Mehdizadeh2017}. In modified gravity theories (MGTs), the stress-energy tensor is replaced with effective stress-energy tensor with higher order curvature terms. MGTs have been utilized to address not only the exotic matter problem but also a number of other challenges in contemporary observational astrophysics and cosmology \cite{Jain2012,Vikram2013,Cabre2012,Planck2015,Amendola2006,Amarzguioui2005}. Due to the lack of observations so far, the geometrical and material properties of WHs, such as the shape function $b(r)$ and equation of state (EoS) still remain unknown. So far, numerous variants of the shape function $b(r)$ have been proposed and analyzed in Refs. \cite{Konoplya2018,Kuhfittig1999,Godani2018}. One of the earliest and simplest adaptations of Einstein-Hilbert action is known as $f(R)$ modified gravity \cite{Buchdahl1970}, where $f(R)$ is a function of the Ricci scalar. This theory has a lot of contributions in the literature that account for the dark energy problem. Starobinsky \cite{Starobinsky1980} discussed the $f(R)$ model in the early $1980$s, using the formula $f(R) = R + \alpha R^2$, where $\alpha > 0$ implies an inflationary universe scenario. This $f(R)$ theory was further developed by many researchers \cite{Thongkool2009,Appleby2009,Nojiri2010}. Many scientists have recently investigated the dynamics of cosmological models utilizing $f(R)$ gravity in a variety of ways \cite{Capozziello2018,Sbisa2018,Elizalde2018,Nascimento2018,Odintsov2019}. Unsurprisingly, these models may contain flaws as well. For example, solar system tests have ruled out the majority of the $f(R)$ models proposed thus far \cite{Chiba2003,Nojiri2007}.\par

So to obtain a better understanding of wormhole solutions, the $f(R)$ theory is extended to the $f(R,\,T)$ gravity theory, where $R$ represents the Ricci scalar and $T$ signifies the trace of the stress-energy tensor. In $2011$, Harko et al. \cite{Harko2011} proposed the $f(R,\,T)$ gravity. This $f(R,\,T)$ gravity has been studied in cosmology \cite{Moraes2015a,Moraes2016,Myrzakulov2012}, thermodynamics \cite{Momeni2015,Harko2014}, and astrophysics of compact objects \cite{Moraes2015,Zubair2015,Shamir2015,Zubair2015a}. $f(R)$ and $f(T)$ theories, in particular, are specific cases of $f(R,\,T)$ gravity. The $f(R,\,T)$ gravity theory combines $f(R)$ and $f(T)$ gravity. Many functional versions of $f(R,\,T)$ theories have been investigated in various contexts for the effect of cosmic dynamics. The split-up $f(R,\,T) = f_1 (R) + f_2 (T )$, where $f_1$ and $f_2$ are arbitrary functions of $R$ and $T$, respectively, has attracted a lot of attention since it allows one to investigate $R$'s contributions without identifying $f_2 (T )$. Likewise, without identifying $f_1 (R)$, one can investigate contributions from $T$. The reconstruction of $f(R,\,T)$ gravity is explored in such separable theories \cite{Houndjo2011}. Several authors have used the formula $f(R,\,T) = f_1 (R) + f_2 (T )$ to analyze cosmic dynamics from various perspectives \cite{Mirza2014,Correa2015,Zaregonbadi2016,Das2016,Yousaf2016} (and further references therein).\par

Many cosmologists have investigated wormhole solutions in modified theories of gravity from various perspectives. In the context of $f(R,\,T)$ gravity, with a constant redshift function $\phi$, a lot of works have been done on different situations of wormhole geometry \cite{Azizi2012,Zubair2016}. Now we are going to discuss some contributions of several cosmologists on wormhole solutions in $f(R,\,T)$ gravity theory. Moraes et al. \cite{Moraes2016a} gave analytical solutions for static wormholes in the $f(R,\,T)$ gravity. In $f(R,\,T)$ gravity, Zubair et al. \cite{Zubair2016} investigated static and spherically symmetric wormholes with various fluids, where $f(R,\,T) = f(R) + \lambda T$. In the context of Gaussian and Lorentzian distributions in non-commutative geometry, Zubair et al. \cite{Zubair2017} proposed a wormhole model in $f(R,\,T)$ gravity. Bhatti et al. \cite{Bhatti2018} used an exponential model to study spherically symmetric wormhole models and find wormhole solutions. Wormhole solutions for dust and non-dust distributions have been investigated in the light of Noether symmetry within a framework of $f(R,\,T)$ gravity in \cite{Sharif2019}. Elizalde and Khurshudyan \cite{Elizalde2019} investigated traversable wormhole solutions in $f(R,\,T)$ gravity by considering the various forms of energy density. Sahoo et.al. \cite{Sahoo2017} studied WH solutions using a modified gravity given by $f(R,\,T) = f(R) + \lambda T$ with a shape function from the Ref. \cite{Heydarzade2014}.

The theoretical construction of wormhole geometries is based on the fact that one must first solve Einstein's field equations by setting the form of the metric potential functions or by applying a precise equation of state that links pressure to energy density. Our specific goal of this paper is to obtain a complete wormhole solution within $f(R,\,T)$ gravity by making an additional assumption that our spacetime permits a one-parameter group of conformal motions, i.e. conformal Killing vectors, which will be discussed in the next section of this paper. This type of systematic approach was previously considered by Boehmer et al. \cite{Boehmer2007,Boehmer2008}. The Einstein field equations provide a natural connection between geometry and matter in the study of conformal symmetry. Kuhfittig recently investigated wormholes that admit a one-parameter group of conformal motions \cite{Kuhfittig2015,Kuhfittig2016}.

The contents of the paper are organized as follows. The basic field equations of $f(R,\,T)$ gravity are briefly reviewed in Sect.~\ref{sec2} with a brief outline of the conformal Killing vectors for a spherically symmetric metric. In Sect.~\ref{sec3} we propose an exact wormhole model by taking into account a specific equation of state. Exterior spacetime and matching conditions have been discussed in Sect.~\ref{sec4}. In Sect.~\ref{sec5}, we will discuss some physical properties, e.g., proper radial distance, active mass function, total Gravitational Energy, energy conditions. Finally, in Sect.~\ref{sec6} we have given some discussion on our obtained model.

\section{$f(R,\,T)$ gravity and basic field equations}\label{sec2}
The basic field equations in the $f(R,\,T)$ theory of gravity will be described here. $f(R,\,T)$ gravity is an extended form of Einstein's general theory of relativity. The Einstein Hilbert action for this gravity is given by,
\begin{eqnarray}\label{action}
S&=&\int \left[\frac{1}{16 \pi} f(R,T)+ \mathcal{L}_m\right]\sqrt{-g} d^4 x,
\end{eqnarray}
where $f(R,\,T )$ is the general function of Ricci scalar $R$ and trace $T$ of the energy-momentum tensor $T_{\mu \nu}$, the Lagrangian matter density is denoted by $\mathcal{L}_m$ with $g = det(g_{\mu \nu}$).\\
The stress-energy tensor of matter $T_{\mu \nu}$ is calculated as,
\begin{eqnarray}\label{tmu1}
T_{\mu \nu}&=&-\frac{2}{\sqrt{-g}}\frac{\delta \sqrt{-g}\mathcal{L}_m}{\delta \sqrt{g_{\mu \nu}}},
\end{eqnarray}
The general field equations for action given in (\ref{action}) is given as,
\begin{widetext}
\begin{eqnarray}\label{frt}
f_R(R,T)R_{\mu \nu}-\frac{1}{2}f(R,T)g_{\mu \nu}+(g_{\mu \nu }
\Box-\nabla_{\mu}\nabla_{\nu})f_R(R,T)&=&8\pi T_{\mu \nu}-
f_T(R,T)T_{\mu \nu}\nonumber\\&&-f_T(R,T)\Theta_{\mu \nu},
\end{eqnarray}
\end{widetext}
with, $f_R(R,T)=\frac{\partial f(R,T)}{\partial
R},~f_T(R,T)=\frac{\partial f(R,T)}{\partial T}$. $\nabla_{\nu}$
represents the covariant derivative associated with the Levi-Civita
connection of $g_{\mu \nu}$, $\Theta_{\mu \nu}=g^{\alpha
\beta}\frac{\delta T_{\alpha \beta}}{\delta g^{\mu \nu}}$ and $\Box
\equiv \frac{1}{\sqrt{-g}}\partial_{\mu}(\sqrt{-g}g^{\mu
\nu}\partial_{\nu})$ represents the D'Alembert operator.\\
The divergence of $T_{\mu \nu}$ is given as \cite{Harko2011,Koivisto2005} ,
\begin{eqnarray}\label{conservation}
\nabla^{\mu}T_{\mu \nu}&=&\frac{f_T(R,T)}{8\pi-f_T(R,T)}
\left[(T_{\mu \nu}+\Theta_{\mu \nu})\nabla^{\mu}\ln f_T(R,T)+\nabla^{\mu}\Theta_{\mu \nu}\right].
\end{eqnarray}
One can check, from eqn.(\ref{conservation}), $\nabla^{\mu}T_{\mu \nu}\neq 0$ for $f_T(R,T)\neq 0$. As a result, the system will not be conserved in the same way that in Einstein gravity it is conserved.\par
The matter Lagrangian density $\mathcal{L}_m$ may be depend on both density and pressure, i.e., $\mathcal{L}_m = \mathcal{L}_m (\rho, p)$, or it only depends on the density of the matter $\rho$ only, so that $\mathcal{L}_m = \mathcal{L}_m (\rho)$ \cite{Harko2011}. In our current paper, we choose the matter Lagrangian density as $\mathcal{L}_m=\rho$  and the expression of $\Theta_{\mu \nu}=-2T_{\mu \nu}-\rho g_{\mu\nu}.$\\
In $f(R,\,T)$ gravity, we consider, $f(R,T)= R+2 \gamma T$, to discuss the coupling effects of matter and curvature components, where $\gamma$ is a dimensionless coupling constant.\par
The interior of a static spherically symmetric spacetime is defined by the following line element in (3+1) dimensions:
\begin{equation}\label{line}
ds^{2}=e^{\nu(r)}dt^{2}-e^{\lambda(r)}dr^{2}-r^{2}(d\theta^2+\sin^2\theta d\phi^2),
\end{equation}
where $\nu(r)=\phi(r)$ and $e^{\lambda(r)}=\left(1-\frac{b(r)}{r}\right)^{-1}$. $\phi(r)$ and $b(r)$ are respectively called the redshift function and shape function of the wormhole and these two functions should satisfy the following conditions:
\begin{itemize}
  \item The wormhole throat joins two asymptotic regions and is placed at the radial coordinate $r_0$, where $b(r_0)= r_0$.
  \item The flaring-out requirement,$\frac{b(r) - rb'(r)}{
2b^2(r)} > 0$, which should valid at or near the throat, must be satisfied by the shape function $b(r)$. This reduces to $b'(r_0)<1$ near the wormhole's throat.
\item The shape function should meet the condition $1-\frac{b(r)}{r}>0$ for the radial coordinates $r>r_0$ in order to maintain the proper signature of the metric.
\item The metric functions must obey the requirements $\phi(r)$ and $b(r)/r$ tend to zero as r approaches to $\infty$ in order to have asymptotically flat geometries. For non-asymptotically flat wormholes, these criteria can obviously be relaxed.
\item It is also necessary for $\phi(r)$ to be finite and nonzero throughout spacetime to avoid horizons and singularities.
\end{itemize}
The constraints listed above give a minimum set of requirements for describing the geometry of two asymptotically flat regions joined by a bridge \cite{Dadhich:2001fu}.\\
For the line element (\ref{line}), the field equations in $f(R,\,T)$ gravity are given by,
\begin{eqnarray}
\kappa\rho^{\text{eff}}&=&\frac{\lambda'}{r}e^{-\lambda}+\frac{1}{r^{2}}(1-e^{-\lambda}),\label{f1}\\
\kappa p_r^{\text{eff}}&=& \frac{1}{r^{2}}(e^{-\lambda}-1)+\frac{\nu'}{r}e^{-\lambda},\label{f2} \\
\kappa p_t^{\text{eff}}&=&\frac{1}{4}e^{-\lambda}\left[2\nu''+\nu'^2-\lambda'\nu'+\frac{2}{r}(\nu'-\lambda')\right], \label{f3}
\end{eqnarray}
where the expression of effective matter density, effective radial and transverse pressure are given by,
\begin{eqnarray}
\rho^{\text{eff}}&=& \rho+\frac{\gamma}{\kappa}( \rho-p_r-2p_t),\label{r1}\\
p_r^{\text{eff}}&=& p_r+\frac{\gamma}{\kappa}(\rho+3p_r+2p_t),\label{r2}\\
p_t^{\text{eff}}&=& p_t+\frac{\gamma}{\kappa}(\rho+p_r+4p_t).\label{r3}
\end{eqnarray}
Here $\rho^{\text{eff}}$, $p_r^{\text{eff}}$ and $p_t^{\text{eff}}$ indicates the matter density, radial and transverse pressure respectively in Einstein's gravity and $\rho,\,p_r$ and $p_t$ respectively denotes the density and pressure in modified gravity.\\
The current work has been approached in a more systematic manner in order to find exact solutions and investigate the inherent relationship between geometry and matter. For example, a well-known idea is that spherically symmetric static spacetime has conformal symmetry under conformal killing vectors (CKV), which can be expressed as,
\begin{equation}\label{con}
\mathcal{L}_\xi  g_{ik}=\xi_{i;k}+\xi_{k;i}=\psi  g_{ik},
\end{equation}
The Lie derivative operator and the conformal factor are denoted by $\mathcal{L}$ and $\psi$, respectively. The conformal symmetry is generated by the vector $\xi$, which induces the metric $g$ to be conformally transferred onto itself along $\xi$. The following expressions are derived from the aforementioned equations:
\begin{eqnarray}
\xi^{1}\nu'=\psi,\,
\xi^{4}=C_1,\,
\xi^{1}=\frac{\psi r}{2},\,
\xi^{1}\lambda'+2\xi^{1},_1=\psi.
\end{eqnarray}
~~~~~Here `prime' and `comma' stand for the derivative and partial derivative with respect to `r' and $C_1$ is a constant.\\
The above equations yield,
\begin{eqnarray}
e^{\nu}&=&C_2^{2}r^{2}, \label{eq11}\\
e^{\lambda}&=&\left(\frac{C_3}{\psi}\right)^{2},\label{eq12}\\
\xi^{i}&=&C_1\delta_{4}^{i}+\left( \frac{\psi r}{2}\right)\delta_1^{i}.\label{eq13}
\end{eqnarray}
Where $C_2$ and $C_3$ are constants of integrations.\par
Using eqns. (\ref{eq11})-(\ref{eq13}), Einstein field equations (\ref{f1})-(\ref{f3}) become :
\begin{eqnarray}
\kappa\rho + \gamma( \rho-p_r-2p_t)&=& f(r),\label{j1} \\
\kappa p_r + \gamma (\rho+3p_r+2p_t)&=& g(r),\label{j2}\\
\kappa p_t +\gamma (\rho+p_r+4p_t)&=& h(r),\label{j3}
  \end{eqnarray}
	where, $\kappa= 8 \pi$ and $f(r),\,g(r)$, $h(r)$ are functions of `r' defined as follows:
\begin{eqnarray*}
  f(r)&=&\frac{1}{r^2}\left[1-\frac{\psi^2}{C_3^2}\right]-\frac{2\psi\psi'}{rC_3^2}, \\
  g(r)&=&\frac{1}{r^2}\left[3\frac{\psi^2}{C_3^2}-1\right],\\
  h(r)&=&\frac{\psi^2}{r^2C_3^2}+\frac{2\psi\psi'}{rC_3^2}.
  \end{eqnarray*}
By solving eqns. (\ref{j1})-(\ref{j3}) we get,
\begin{eqnarray}
\rho &=& \frac{(\kappa + 5 \gamma)f(r)+\gamma g(r) +2 \gamma h(r)}{(\kappa + 2 \gamma)(\kappa + 4 \gamma)}, \label{eq14}\\
p_r &=& \frac{-\gamma f(r) + (\kappa + 3 \gamma)g(r) - 2 \gamma h(r)}{(\kappa + 2 \gamma)(\kappa + 4 \gamma)}, \label{eq15} \\
p_t &=& \frac{-\gamma f(r)-\gamma g(r)+ (\kappa + 2 \gamma)h(r) }{(\kappa + 2 \gamma)(\kappa + 4 \gamma)}. \label{eq16}
\end{eqnarray}
	
\section{Proposed model of wormhole}\label{sec3}
\begin{figure}[htbp]
    \centering
        \includegraphics[scale=.4]{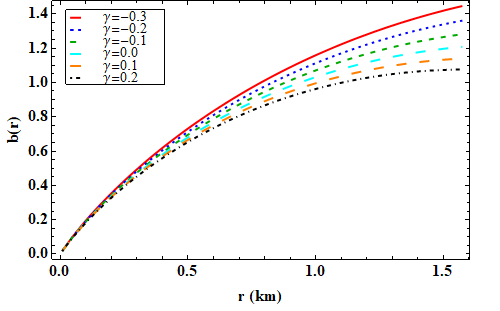}
        \includegraphics[scale=.4]{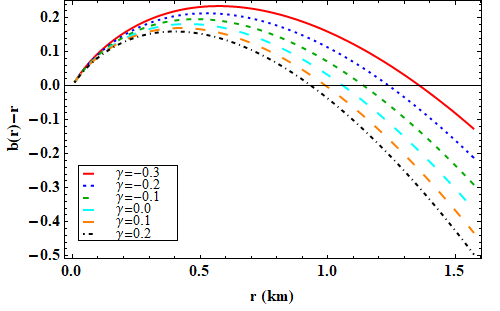}
         \includegraphics[scale=.4]{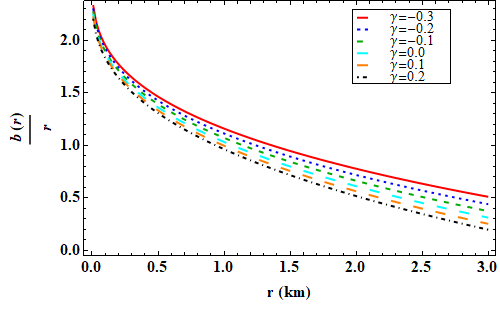}
          \includegraphics[scale=.4]{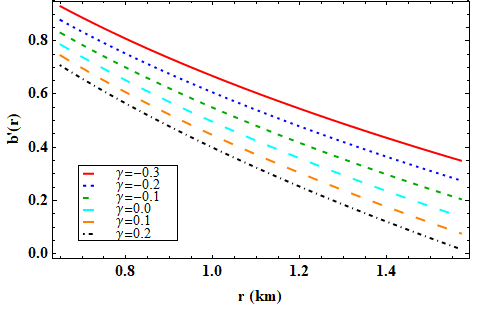}
       \caption{The shape function $b(r)$, throat of the wormhole, $b(r)/r$ and $b'(r)$ are shown against r. \label{fig1}}
\end{figure}

According to Morris and Throne \cite{Morris1988}, developing a wormhole solution necessitates the use of an unusual form of matter known as `exotic matter', which is a necessary component for a traversable wormhole to exist. The energy density, $\rho$, of such matter is positive, but the radial pressure, $p_r$, must be negative. Recent theoretical advances reveal that the expansion of our current universe is speeding up, and dark energy is a suitable candidate to explain it. In this context, we look at how to obtain traversable wormholes utilizing the phantom energy equation of state which is given by the following relationship:
\begin{eqnarray}\label{j4}
p_r=\omega \rho,~~~ \text{with}~~\omega<-1,
\end{eqnarray}
Using (\ref{eq14}) and (\ref{eq15}), from relation (\ref{j4}) we get,
\begin{widetext}
\begin{eqnarray}
  \big(8 \gamma+\kappa (3 + \omega)\big)\frac{\psi^2}{C_3^2}-\big(\gamma - (3 \gamma + \kappa) \omega\big)\frac{2r\psi\psi'}{C_3^2} &=& (4 \gamma + \kappa) (1 + \omega),
\end{eqnarray}
\end{widetext}
Solving the above equation one can obtain,
\begin{eqnarray}
\frac{\psi^2}{C_3^2}&=&Dr^{\frac{8 \gamma + \kappa (3 + \omega)}{\gamma - (3 \gamma + \kappa) \omega}}+\frac{(4 \gamma + \kappa) (1 + \omega)}{8 \gamma + \kappa (3 + \omega)},
\end{eqnarray}
where $D$ is a dimensionless constant of integration.\\
The expressions for matter density, radial, and transverse pressure are as follows:
\begin{eqnarray}
\rho&=&\frac{2 (4 \gamma + \kappa)}{(2 \gamma + \kappa) (8 \gamma + \kappa (3 + \omega)) r^2}+\frac{3 D r^{-\frac{3 (2 \gamma + \kappa) (1 + \omega)}{
  \kappa \omega + \gamma (-1 + 3 \omega)}}}{\kappa \omega + \gamma (-1 + 3 \omega)},
  \\
p_r&=&\frac{2\omega (4 \gamma + \kappa)}{(2 \gamma + \kappa) (8 \gamma + \kappa (3 + \omega)) r^2}+\frac{3 D\omega r^{-\frac{3 (2 \gamma + \kappa) (1 + \omega)}{
  \kappa \omega + \gamma (-1 + 3 \omega)}}}{\kappa \omega + \gamma (-1 + 3 \omega)},
  \\
  p_t&=&\frac{\kappa (1 + \omega)}{(2 \gamma + \kappa) (8 \gamma + \kappa (3 + \omega)) r^2}-\frac{3 D r^{-\frac{3 (2 \gamma + \kappa) (1 + \omega)}{
  \kappa \omega + \gamma (-1 + 3 \omega)}}}{\kappa \omega + \gamma (-1 + 3 \omega)}.
\end{eqnarray}
Now using Eqn. (\ref{eq12}), the expression of metric potential is obtained as,
\[e^{-\lambda}=Dr^{\frac{8 \gamma + \kappa (3 + \omega)}{\gamma - (3 \gamma + \kappa) \omega}}+\frac{(4 \gamma + \kappa) (1 + \omega)}{8 \gamma + \kappa (3 + \omega)},\]
Now using the relationship $e^{\lambda}=\frac{1}{1-b(r)/r}$, the shape function of the present model of wormhole is obtained as,
\[b(r)=\frac{2 (\kappa - 2 \gamma (-1 + \omega))}{8 \gamma + \kappa (3 + \omega)}r-Dr^{\frac{3 (\kappa - \gamma (-3 + \omega))}{\gamma - (3 \gamma + \kappa) \omega}}.\]
Fig.~\ref{fig1} shows the graphical behaviour of the $b(r)$, $b(r)-r$, $b(r)/r$, and $b'(r)$ for wormhole for different values of $\gamma$ with $D=1.5$ and $\omega=-2.21$. $b(r)-r$ cuts the r-axis at the throat at $r=r_0$ as shown in the figure. In Table~\ref{t1} we have shown the position of the throat of the wormhole for different values of $\gamma$. In Fig.~\ref{fig1}, we also see that $b'(r)<1$, which obeys the flaring out requirement. Furthermore, as shown in the figure, the asymptotic behaviour of $b(r)/r$ does not approach to zero as `r' approaches to $\infty$, the redshift function also does not approach zero as r approaches to $\infty$, indicating that the wormhole spacetime is not asymptotically flat.

\begin{table}[t]
\centering
\caption{Position of the throat $r_0$ for different values of $\gamma$}\label{t1}
\begin{tabular}{@{}ccccccccccccc@{}}
\hline
$\gamma$&& $r_0$\\
\hline
-0.3 &&    1.36\\
-0.2 &&    1.24\\
-0.1 &&   1.14 \\
0.0 &&   1.06 \\
0.1 &&  0.99\\
0.2 &&  0.935\\
\hline
\end{tabular}
\end{table}

\begin{figure}[htbp]
    \centering
        \includegraphics[scale=.2]{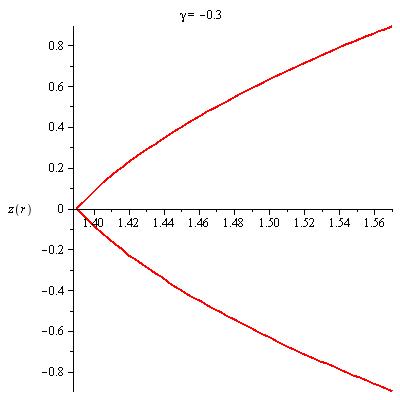}
		\includegraphics[scale=.2]{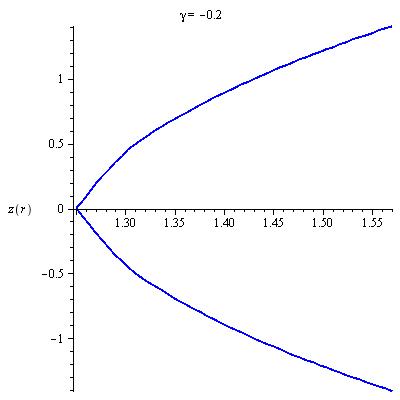}
        \includegraphics[scale=.2]{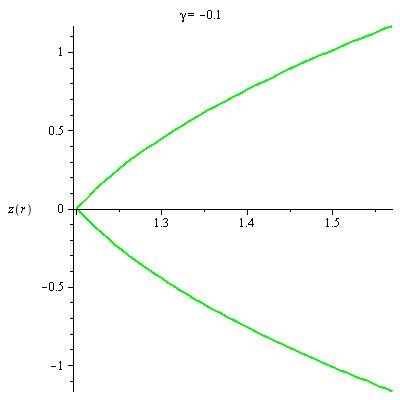}
        \includegraphics[scale=.2]{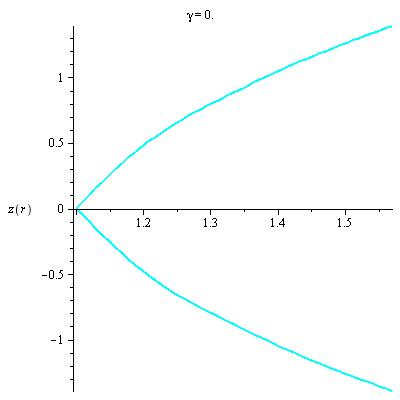}
        \includegraphics[scale=.2]{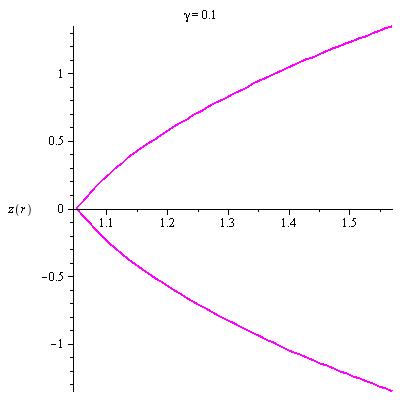}
         \includegraphics[scale=.2]{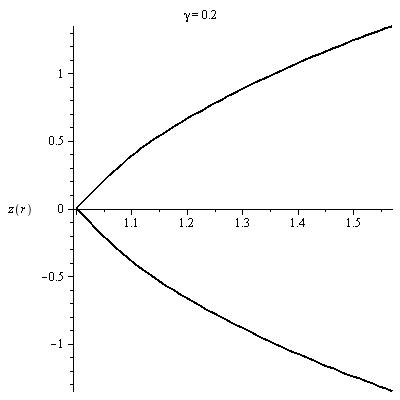}
        \caption{The embedding diagram of wormholes for different values of $\gamma$ \label{z11}}
\end{figure}

\begin{figure}[htbp]
    \centering
    \includegraphics[scale=.3]{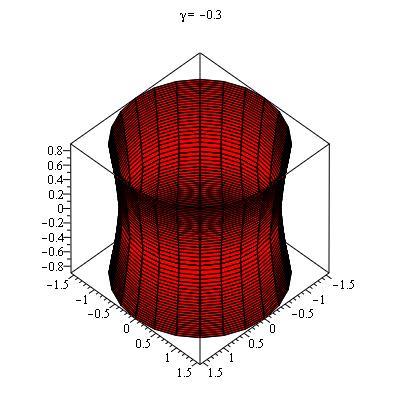}
    \includegraphics[scale=.3]{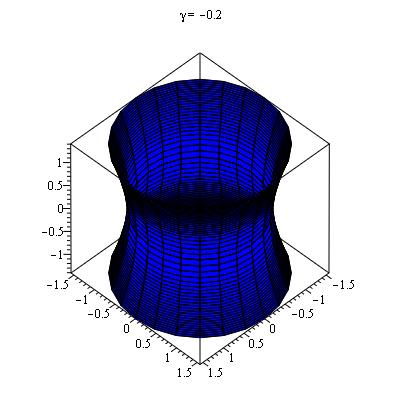}
    \includegraphics[scale=.3]{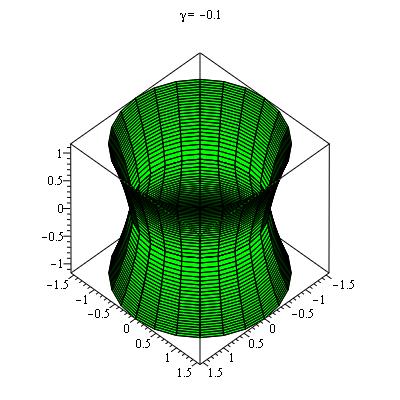}
    \includegraphics[scale=.3]{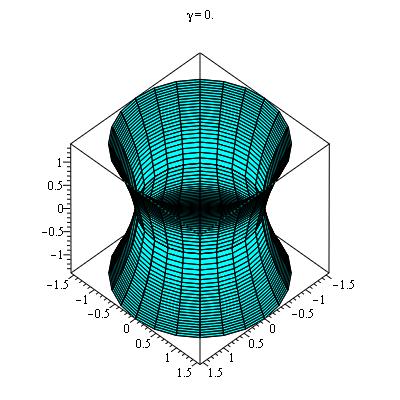}
    \includegraphics[scale=.3]{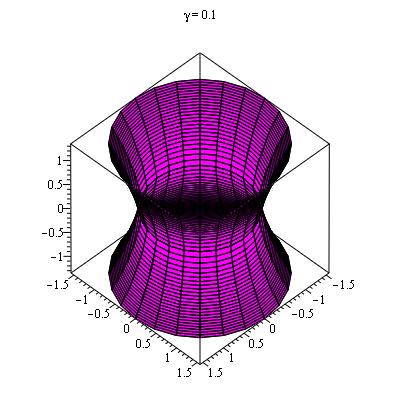}
    \includegraphics[scale=.3]{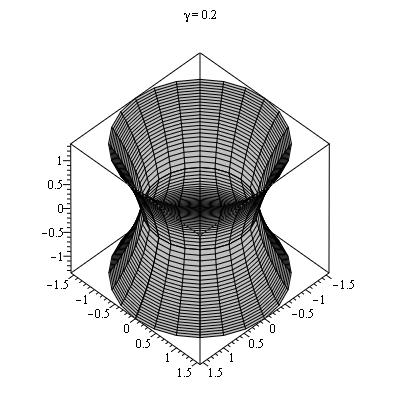}
        \caption{The full visualization of wormholes for different values of $\gamma$.\label{w11}}
\end{figure}

Following Morris and throne \cite{Morris1988}, the function $z(r)$ represents the wormhole's embedding surface
which is the solution of following differential equation:
\begin{equation}\label{z}
\frac{dz}{dr}=\pm\frac{1}{\sqrt{\frac{r}{b(r)}-1}}
\end{equation}
The above differential equation diverges at the throat of the
wormhole and therefore it concludes that the embedding surface is vertical at the throat.\\
Eqn. (\ref{z}) gives,
\begin{equation}\label{z1}
z=\pm\int_{r_0^{+}}^{r}\frac{dr}{\sqrt{\frac{r}{b(r)}-1}},
\end{equation}
here, $r_0$ is the throat of the wormhole.\\
The integral in equation (\ref{z1}) cannot be solved analytically. However, we can do it numerically. The numerical values are obtained by setting a specific lower limit value outside the wormhole throat and changing the upper limit, as shown in Tables \ref{table1}-\ref{table6}, for $\gamma=-0.3,\,-0.2,\,-0.1,\,0,\,0.1$ and $0.2$ respectively. Fig.~\ref{z11} shows the embedding diagram $z(r)$ of wormholes for various values of the coupling constant $\gamma$. Fig.~\ref{w11} shows the entire visualization of the wormholes for different values of $\gamma$ mentioned in the figures produced by rotating Fig.~\ref{z11} around the `z-axis'.
\begin{table}[t]
\centering
\caption{Values of $z(r)$, $l(r)$ and $E_g$ for different $r$ by taking $r_0^+ = 1.39$, $D=1.5$, $\omega = -2.21$ and $\gamma=-0.3$.}\label{table1}
\begin{tabular}{@{}ccccccccccccc@{}}
\hline
$r$ & $z(r)$ & $l(r)$ & $E_g$\\
\hline
1.41 & 0.158945 &0.160205& 0.678778 \\
1.43& 0.287932 &0.290737 & 0.677838\\
1.45&0.399285 &0.403874 &0.677067\\
1.49&0.589172 &0.597939 &0.675841\\
1.51&0.67282 &0.683947 &0.675337\\
1.53&0.750924 &0.764571 & 0.674887\\
1.57&0.894034 & 0.913171&0.674114\\
\hline
\end{tabular}
\end{table}

\begin{table}[t]
\centering
\caption{Values of $z(r)$, $l(r)$ and $E_g$ for different $r$ by taking $r_0^+ = 1.25$, $D=1.5$, $\omega = -2.21$ and $\gamma=-0.2$.}\label{table2}
\begin{tabular}{@{}ccccccccccccc@{}}
\hline
$r$ & $z(r)$ & $l(r)$ & $E_g$ \\
\hline
1.3&0.43116 &0.434211&0.616648 \\
1.35&0.689857&0.697731&0.614918\\
1.4&0.893041&0.906994&0.613721\\
1.45&1.06552&1.08658&0.61282\\
1.5&1.21771&1.24678&0.612114 \\
1.57&1.40683&1.44845&0.611356\\
\hline
\end{tabular}
\end{table}

\begin{table}[t]
\centering
\caption{Values of $z(r)$, $l(r)$ and $E_g$ for different $r$ by taking $r_0^+ = 1.2$, $D=1.5$, $\omega = -2.21$ and $\gamma=-0.1$.}\label{table3}
\begin{tabular}{@{}ccccccccccccc@{}}
\hline
$r$ & $z(r)$ & $l(r)$ & $E_g$  \\
\hline
1.25&0.250562&0.25554&0.568355\\
1.3&0.446391&0.45767&0.567233\\
1.35&0.612213&0.630877&0.566399\\
1.4&0.758284&0.785276&0.565753\\
1.45&0.890038&0.926204&0.565243\\
1.5&1.01077&1.05689&0.564836\\
1.57&1.1653&1.22654&0.564398\\
\hline
\end{tabular}
\end{table}

\begin{table}[t]
\centering
\caption{Values of $z(r)$, $l(r)$ and $E_g$ for different $r$ by taking $r_0^+ = 1.1$, $D=1.5$, $\omega = -2.21$ and $\gamma=0.0$.}\label{table4}
\begin{tabular}{@{}ccccccccccccc@{}}
\hline
$r$ & $z(r)$ & $l(r)$ & $E_g$ \\
\hline
1.2&0.480638&0.491243&0.526922\\
1.3&0.795979&0.82215&0.525413\\
1.4&1.04666&1.09209&0.524513\\
1.5&1.25954&1.32731&0.523956\\
1.57&1.39274&1.47779&0.523701\\
\hline
\end{tabular}
\end{table}

\begin{table}[t]
\centering
\caption{Values of $z(r)$, $l(r)$ and $E_g$ for different $r$ by taking $r_0^+ = 1.05$, $D=1.5$, $\omega = -2.21$ and $\gamma=0.1$.}\label{table5}
\begin{tabular}{@{}ccccccccccccc@{}}
\hline
$r$ & $z(r)$ & $l(r)$ & $E_g$ \\
\hline
1.1&0.234286&0.239603&0.493524\\
1.2&0.570816&0.590802&0.491819\\
1.3&0.828309&0.867089&0.490856\\
1.4&1.04317&1.10412&0.490279\\
1.5&1.22998&1.31603&0.489941\\
1.57&1.34827&1.45349&0.489803\\
\hline
\end{tabular}
\end{table}

\begin{table}[t]
\centering
\caption{Values of $z(r)$, $l(r)$ and $E_g$ for different $r$ by taking $r_0^+ = 1$, $D=1.5$, $\omega = -2.21$ and $\gamma=0.2$.}\label{table6}
\begin{tabular}{@{}ccccccccccccc@{}}
\hline
$r$ & $z(r)$ & $l(r)$ & $E_g$ \\
\hline
1.1&0.388527&0.401406&0.465312\\
1.2&0.663385&0.69397&0.46419\\
1.3&0.886099&0.938147&0.463542\\
1.4&1.07665&1.15337&0.46317\\
1.5&1.24455&1.34882&0.462977\\
1.57&1.35162&1.47674&0.462917\\
\hline
\end{tabular}
\end{table}

\section{Exterior spacetime and smooth matching }\label{sec4}
Since the wormhole spacetime is non-asymptotically flat, it should match the exterior Schwarzschild spacetime given by the following metric at some junction radius $r = R$ :
\begin{equation}
ds^{2}=\left(1-\frac{2m}{r}\right)dt^{2}-\frac{dr^{2}}{1-\frac{2m}{r}}-r^{2}\left(d\theta^{2}+\sin^{2}\theta d\phi^{2}\right),
\end{equation}
The matching takes place at a radius greater than the event horizon, which yields,
\begin{eqnarray}
C_2^2R^2=1-\frac{2M}{R},\,
1-\frac{2M}{R}=1-\frac{b(R)}{R},
\end{eqnarray}
where, $M=m(R)$. Solving the above two equations we get,
\begin{eqnarray}
M&=&\frac{(\kappa - 2 \gamma (-1 + \omega))}{8 \gamma + \kappa (3 + \omega)}R-\frac{D}{2}R^{\frac{3 (\kappa - \gamma (-3 + \omega))}{\gamma - (3 \gamma + \kappa) \omega}},\\
C_2^2&=&\frac{1}{R^2}\left[DR^{\frac{8 \gamma + \kappa (3 + \omega)}{\gamma - (3 \gamma + \kappa) \omega}}+\frac{(4 \gamma + \kappa) (1 + \omega)}{8 \gamma + \kappa (3 + \omega)}\right].
\end{eqnarray}
\section{Some physical properties}\label{sec5}
In the context of the $f(R,\,T)$ theory of gravitation, we have examined at Morris-Thorne wormholes, which are static and spherically symmetric traversable wormholes. We basically describe here wormhole solutions in a viable $f(R,T)$ gravity with the form $f(R,\,T)=R+2 \gamma T$. This model was chosen over the other  alternatives because it is the most straightforward in terms of numerical computation. Another benefit of this model is that the reported results are easy to compare or distinguish from the GR counterparts by simply taking the limit $\gamma\rightarrow 0$. In this section we are going to verify some physical properties of our proposed model.
\subsection{Proper radial distance}
\begin{figure}[htbp]
    \centering
    \includegraphics[scale=.4]{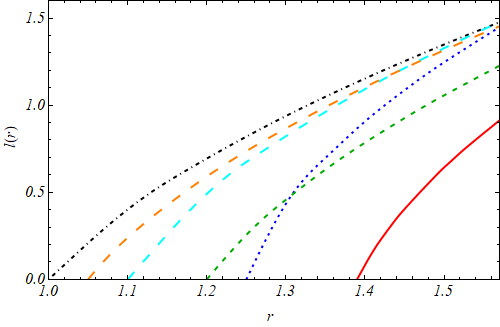}
        \caption{Proper radial distance is shown against `r' for different values of $\gamma$. The description of the curve is same as Fig.~\ref{fig1} \label{lr}}
\end{figure}

The proper radial distance of the wormhole is given by,
\begin{equation}\label{l}
l(r)=\pm\int_{r_0^{+}}^{r}\frac{dr}{\sqrt{1-\frac{b(r)}{r}}}.
\end{equation}
With the help of numerical integration we have performed the above integral for different values of $\gamma$ and the results are presented in Tables \ref{table1}-\ref{table6}. The profiles of $l(r)$ for different values of $\gamma$ are shown in Fig.~\ref{lr}.
\subsection{Active mass function}
The active mass function for our wormhole is calculated as,
\begin{widetext}
\begin{eqnarray}
M_{active}&=&\int_{r_0^+}^{r}4\pi\rho r^{2}dr,\nonumber\\
&=&\frac{\kappa}{2} \left[\frac{
   2 (4 \gamma + \kappa) r}{(2 \gamma + \kappa) (8 \gamma +
      \kappa (3 + \omega))} + \frac{
   D r^{\frac{3 (\kappa - \gamma (-3 + \omega))}{
    \gamma - (3 \gamma + \kappa) \omega}}}{-\kappa + \gamma (-3 + \omega)}\right]_{r_0^+}^r.
\end{eqnarray}
\end{widetext}
where $r_0^+$ indicates outside the throat.\\
\begin{figure}[htbp]
    \centering
        \includegraphics[scale=.38]{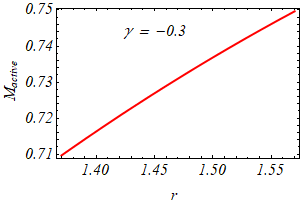}
        \includegraphics[scale=.38]{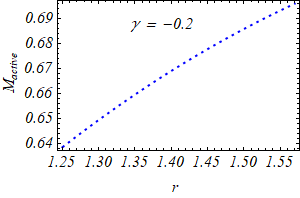}
         \includegraphics[scale=.38]{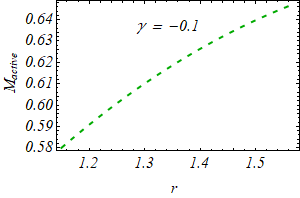}
          \includegraphics[scale=.38]{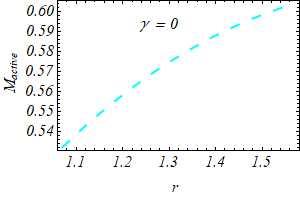}
             \includegraphics[scale=.38]{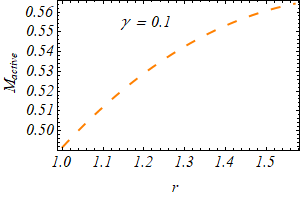}
                \includegraphics[scale=.38]{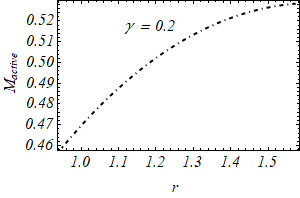}
       \caption{Active mass function of the wormhole has been shown for different values of $\gamma$. \label{mactive1}}
\end{figure}
In Fig.~\ref{mactive1}, the profile of active mass function of the wormhole is displayed for different values of $\gamma$. $M_{active}$ is positive outside the wormhole throat, and it is a monotonic increasing function of `r', as seen in the figure.

\subsection{Total Gravitational Energy}
\begin{figure}[htbp]
    \centering
        \includegraphics[scale=.38]{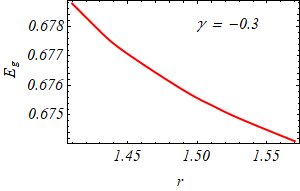}
        \includegraphics[scale=.38]{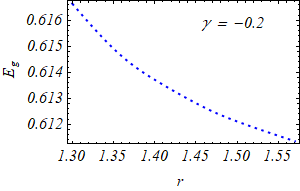}
         \includegraphics[scale=.38]{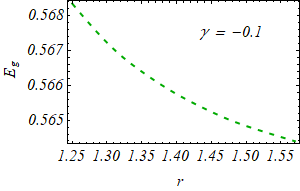}
          \includegraphics[scale=.38]{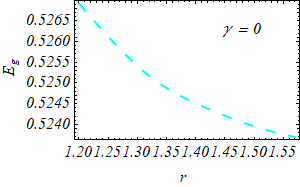}
             \includegraphics[scale=.38]{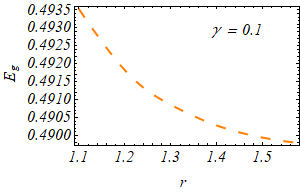}
                \includegraphics[scale=.38]{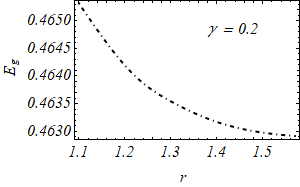}
       \caption{Total gravitational energy of the wormhole has been shown for different values of $\gamma$. \label{eg1}}
\end{figure}

We want to determine the total gravitational energy ($E_g$) of the wormholes using the Lyndell-Bell et al. \cite{Lynden-Bell:2007gof}, Katz et al. \cite{Katz:2006uw} and Nandi et al.\cite{Nandi:2008ij} concepts, which could be determined from the following formula:
\begin{equation}\label{eg}
E_g=Mc^{2}-E_M.
\end{equation}
Where \begin{equation}\label{em}
Mc^{2}=\frac{1}{2}\int_{r_0^{+}}^{R}\rho r^{2}dr+\frac{r_0}{2},\, E_M=\frac{1}{2}\int_{r_0^{+}}^{r}\sqrt{g_{rr}}\rho r^{2}dr,
\end{equation}
where $Mc^{2}$ and $E_M$ are respectively called the total energy and total mechanical energy. From (\ref{em}) and (\ref{eg}) we get,
\begin{equation}
E_g=\frac{1}{2}\int_{r_0^{+}}^{r}[1-(g_{rr})^{\frac{1}{2}}]\rho r^{2} dr
 +\frac{r_0}{2},
 \end{equation}\label{Eg}
 where $g_{rr}=\left(1-\frac{b(r)}{r}\right)^{-1}$.\\
We numerically performed the above integral for different values of $\gamma$ and presented our results in Tables \ref{table1}-\ref{table6} due to the complexity of the computation. The profiles of $E_g$ for different values of $\gamma$ is shown in Fig.~\ref{eg1}.

\subsection{Energy conditions}
\begin{figure*}[htbp]
    \centering
         \includegraphics[scale=.4]{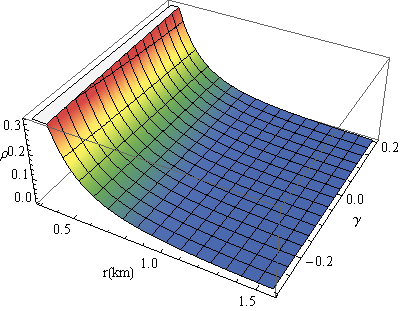}
		\includegraphics[scale=.4]{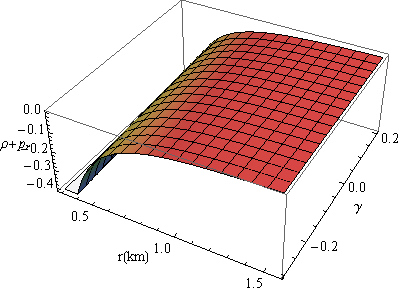}
        \includegraphics[scale=.4]{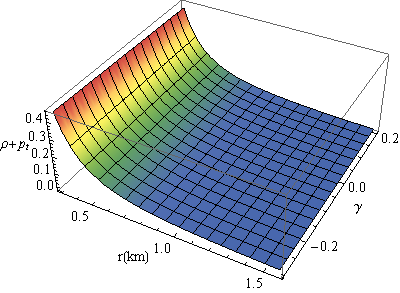}
        \includegraphics[scale=.4]{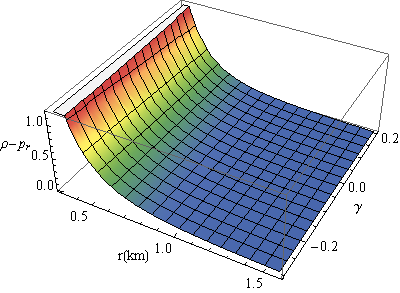}
        \includegraphics[scale=.4]{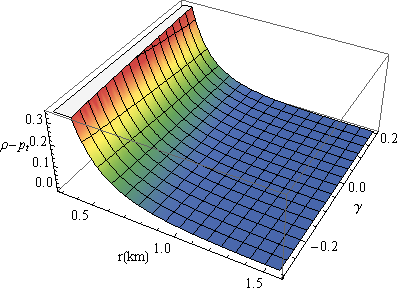}
        \includegraphics[scale=.4]{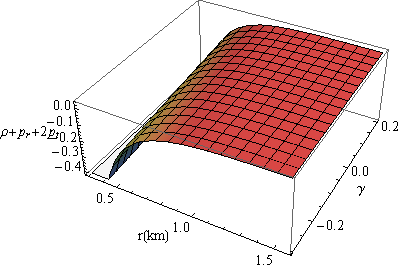}
        \caption{All the energy conditions are shown against `r'.\label{ec}}
\end{figure*}

The energy conditions are utilized in a variety of contexts to obtain a number of results that are applicable to different situations. The notion of energy conditions arises from the Raychaudhuri equation are given by,
\begin{eqnarray}
\frac{d\theta}{d\tau}&=&-\frac{1}{3}\theta^2-\sigma_{\mu \nu}\sigma^{\mu \nu}+\omega_{\mu \nu}\omega^{\mu \nu}-R_{\mu \nu}u^{\mu}u^{\nu},
\end{eqnarray}
where $u^{\mu}$ is a vector field representing the congruence of
`timelike geodesics'. Also $R_{\mu \nu}$, $\theta$, $\theta_{\mu \nu}$ and $\omega_{\mu \nu}$ represent
Ricci tensor, the expansion parameter, the shear and the rotation associated with the congruence, respectively. For the case of a congruence of `null geodesics' defined by the vector field $k^{\mu}$, the Raychaudhuri equation becomes
\begin{eqnarray}
\frac{d\theta}{d\tau}&=&-\frac{1}{2}\theta^2-\sigma_{\mu \nu}\sigma^{\mu \nu}+\omega_{\mu \nu}\omega^{\mu \nu}-R_{\mu \nu}k^{\mu}k^{\nu}.
\end{eqnarray}
From both Raychaudhuri equations, it is clear that these energy conditions are independent of any gravity theory and these are purely geometrical.
In both the theories (general relativity and modified gravity theories) energy constraints are extremely significant. There are four different forms of energy constraints, each of which is expressed using well-known geometrical results. These are labeled as weak energy condition (WEC), Dominant energy condition (DEC), Null energy condition (NEC), and Strong energy condition (SEC). These are defined as follows for an anisotropic fluid:
\begin{itemize}
  \item NEC: $\rho + p_r \geq 0$,\, $\rho + p_t \geq 0$,
  \item WEC: $\rho\geq 0$,\, $\rho + p_r \geq 0$,\, $\rho + p_t \geq 0$,
  \item SEC: $\rho + p_r \geq 0$,\, $\rho + p_t \geq 0$,\, $\rho + p_r + 2p_t \geq 0$,
  \item DEC: $\rho > |p_r|,\, \rho > |p_t|$.
\end{itemize}
To check the following inequalities we need the following expressions:
\begin{widetext}
\begin{eqnarray}
  \rho+p_r &=& (1+\omega)\Big[\frac{2 (4 \gamma + \kappa)}{(2 \gamma + \kappa) (8 \gamma + \kappa (3 + \omega)) r^2}+\frac{3 D r^{-\frac{3 (2 \gamma + \kappa) (1 + \omega)}{
  \kappa \omega + \gamma (-1 + 3 \omega)}}}{\kappa \omega + \gamma (-1 + 3 \omega)}\Big],\\
  \rho+p_t&=&\frac{1}{(2 \gamma + \kappa) r^2},\\
  \rho+p_r+2p_t&=&\frac{4 (1 + \omega)}{(8 \gamma + \kappa (3 + \omega)) r^2} + \frac{
 3 D (-1 + \omega) r^{-\frac{3 (2 \gamma + \kappa) (1 + \omega)}{
   \kappa \omega + \gamma (-1 + 3 \omega)}}}{
 \kappa \omega + \gamma (-1 + 3 \omega)},\\
  \rho-p_r &=&  (1-\omega)\Big[\frac{2 (4 \gamma + \kappa)}{(2 \gamma + \kappa) (8 \gamma + \kappa (3 + \omega)) r^2}+\frac{3 D r^{-\frac{3 (2 \gamma + \kappa) (1 + \omega)}{
  \kappa \omega + \gamma (-1 + 3 \omega)}}}{\kappa \omega + \gamma (-1 + 3 \omega)}\Big],\\
  \rho-p_t&=&\frac{8 \gamma + \kappa - \kappa \omega}{(2 \gamma + \kappa) (8 \gamma + \kappa (3 + \omega)) r^2}+\frac{6 D r^{-\frac{3 (2 \gamma + \kappa) (1 + \omega)}{
  \kappa \omega + \gamma (-1 + 3 \omega)}}}{\kappa \omega + \gamma (-1 + 3 \omega)}.
\end{eqnarray}
\end{widetext}
In Fig.~\ref{ec}, we have plotted all the energy conditions. From the figure, one can note our proposed model only satisfies DEC, the rest three energy conditions are violated.

\subsection{TOV equation}
\begin{figure}[htbp]
    \centering
        \includegraphics[scale=.25]{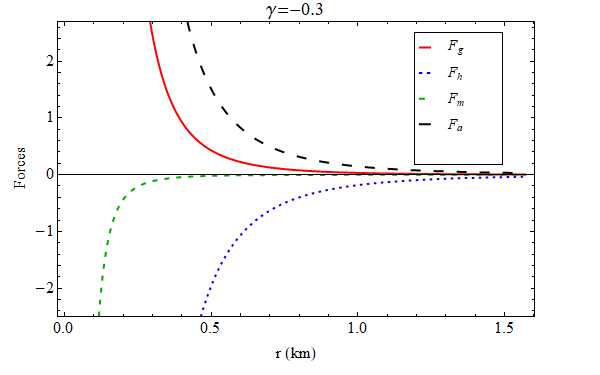}
        \includegraphics[scale=.25]{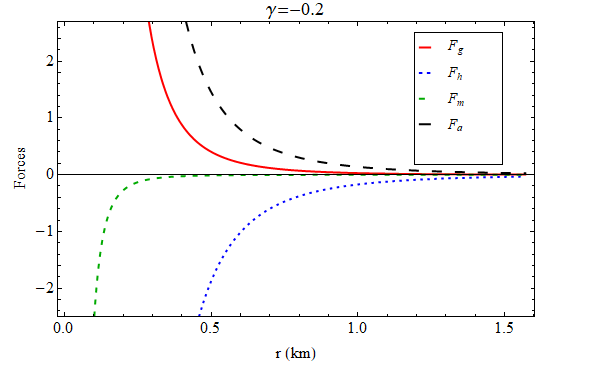}
         \includegraphics[scale=.25]{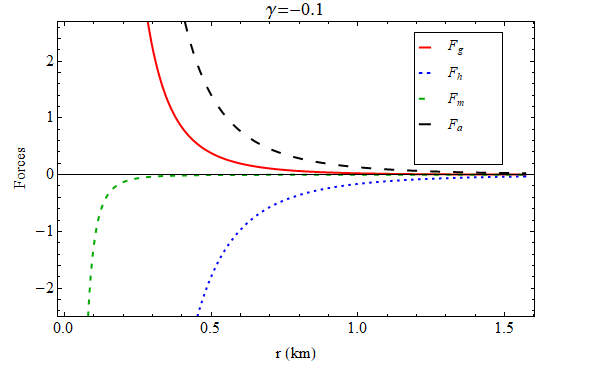}
          \includegraphics[scale=.25]{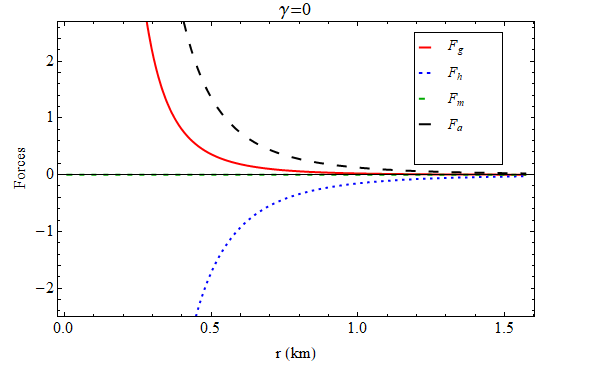}
          \includegraphics[scale=.25]{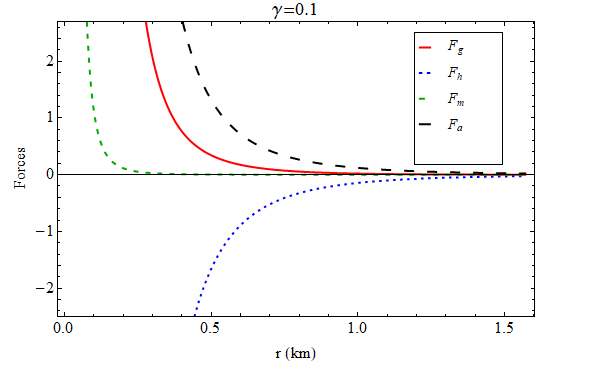}
          \includegraphics[scale=.25]{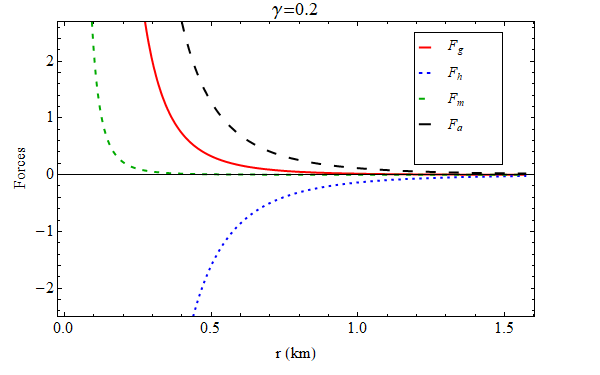}
       \caption{The different forces acting on the system are shown against r for different values of $\gamma$. \label{tov48}}
\end{figure}

{\bf The effect of various forces acting on our current system will be discussed in this subsection. The modified Tolman-Oppeneheimer-Volkoff (TOV) equation in $f(R,\,T)$ gravity can be used to investigate the hydrostatic equilibrium under various forces which is described by the following equation \cite{Bhar:2020abv}:
\[-\frac{\nu'}{2}(\rho+p_r)-\frac{dp_r}{dr}+\frac{2}{r}(p_t-p_r)+\frac{\gamma}{8\pi+2\gamma}(\rho'+p_r'+2p_t')=0\]
In order to attain this hydrostatic balance, the equilibrium equation can be broken down into four separate forces: hydrostatic ($F_h$), gravitational ($F_g$), anisotropic forces ($F_a$) and finally the force due to modified gravity ($F_m$). Furthermore, the explicit forms of these forces are as follows:
\begin{eqnarray}
F_h&=&-\frac{dp_r}{dr}=\omega \left[\frac{4 (4 \gamma + \kappa)}{(2 \gamma + \kappa) \{8 \gamma + \kappa (3 + \omega)\} r^3}+\frac{9 D(2 \gamma + \kappa) (1 + \omega)r^{1 - \frac{3 (2 \gamma + \kappa) (1 + \omega)}{
 \kappa \omega + \gamma (-1 + 3 \omega)}}}{\{\kappa \omega + \gamma (-1 + 3 \omega)\}^2}\right],\\
F_a&=&\frac{2}{r}(p_t-p_r)=\frac{2 \kappa - 2 (8 \gamma + \kappa) \omega}{(2 \gamma + \kappa) (8 \gamma + \kappa (3 + \omega)) r^3}-\frac{6 D(1 + \omega)r^{1 - \frac{3 (2 \gamma + \kappa) (1 + \omega)}{
 \kappa \omega + \gamma (-1 + 3 \omega)}}}{\kappa \omega + \gamma (-1 + 3 \omega)},\\
F_g&=&-\frac{\nu'}{2}(\rho+p_r)=-\frac{1+\omega}{r}\Big[\frac{2 (4 \gamma + \kappa)}{(2 \gamma + \kappa) (8 \gamma + \kappa (3 + \omega)) r^2}+\frac{3 D r^{-\frac{3 (2 \gamma + \kappa) (1 + \omega)}{
  \kappa \omega + \gamma (-1 + 3 \omega)}}}{\kappa \omega + \gamma (-1 + 3 \omega)}\Big],\\
F_m&=&\frac{\gamma}{8\pi+2\gamma}(\rho'+p_r'+2p_t')=-\frac{\gamma}{8\pi+2\gamma}\frac{(1 + \omega) r^{-\frac{3 (\gamma + \kappa + 5 \gamma \omega + 2 \kappa \omega)}{
  \kappa \omega + \gamma (-1 + 3 \omega)}}}{\{8 \gamma + \kappa (3 + \omega)\} \{\kappa \omega + \gamma (-1 + 3 \omega)\}^2}\Big[9D (2 \gamma + \kappa) (-1 + \omega) \times\nonumber\\ && \{8 \gamma +
    \kappa (3 + \omega)\} r^2 +
 8 \{\kappa \omega + \gamma (-1 + 3 \omega)\}^2 r^{\frac{
  3 (2 \gamma + \kappa) (1 + \omega)}{
  \kappa \omega + \gamma (-1 + 3 \omega)}}\Big].
 \end{eqnarray}

It is important to note that if $\gamma=0$, the well-known TOV equation in Einstein gravity can be recovered.\\
The system is in equilibrium under the said forces, as shown in Fig. \ref{tov48}. As seen from the figure, the
anisotropic force and gravitational force are repulsive in nature, on the other hand the hydrostatic force is always attractive in nature irrespective of the value of the coupling constant $\gamma$. The force $F_m$ due to modified gravity is attractive when $\gamma<0$ and repulsive when $\gamma>0$. In the first three figures, the combine effect of gravitational and anisotropic forces are counterbalanced by the joint action of the hydrostatic and modified gravity forces. Moreover, in case of positive $\gamma$, the hydrostatics force is dominating over other three acting forces which is showing that local $F_h$ has a majority role in the balance of the wormhole model for positive values of $\gamma$.}

\section{Discussion}\label{sec6}
In this manuscript, we have examined the wormhole configuration in the background of $f(R,\,T)$ gravity. The accelerated expansion problem of the universe motivated us to go beyond the standard description of gravity. As a result, several modified theories proposed and $f(R,\,T)$ gravity theories have been widely used in modern cosmology without mentioning dark energy for the last decade. It is worth mentioning here that the limit $f(R,\,T)~\rightarrow~R$ recovered the general relativistic bound. However, the interesting motivation about this theory is that the coupling between matter and geometry helps us to resolve some modern cosmology problems.
In this article, we have discussed the wormhole geometries in the context of $f(R,\,T)$ gravity. Specifically, the wormhole configuration possesses a strange matter at its throat to be a traversable wormhole, including a stress-energy tensor that violates the null energy condition. For this configuration, we have developed the wormhole geometries by presuming a static spherically symmetric spacetime as an interior solution with conformal symmetry. The conformal killing vector field describes the spacetime's conformal symmetric properties and reduces the number of unknown quantities. We have examined various types of wormhole solutions and their properties to present a traversable wormhole with asymptotically flat geometries.
For this study, we presumed the phantom energy equation of state i.e., $p_r=\omega \rho,$ with $\omega <-1$ and derived the solution for shape function $b(r)$. The profiles of shape function with $b(r)-r, \,\ b(r)/r,  \,\ b'(r)$ are depicted for different values of $\gamma$ in Fig.~\ref{fig1}. From that Figure, we observed that our obtained shape function successfully cleared all the tests to present a traversable wormhole. The wormhole throat radius for different values of $\gamma$ is presented in Table~\ref{t1}. Furthermore, the 2D and 3D embedding diagrams of the wormholes are illustrated in Fig.~\ref{z11} and \ref{w11}, respectively.
In addition, we have discussed some physical properties of our wormhole solutions. The proper radial distance $l(r)$ has a great property to detect the singularities and help us to avoid the horizon in spacetime. The profiles of $l(r)$ for different values of $\gamma$ are shown in Fig.~\ref{lr}. It is seen that the proper radial distance shows an increasing behavior with finite values throughout the configuration. From Fig.~\ref{mactive1}, it is observed that the active mass function is positive for different values of $\gamma$.
Moreover, the gravitation potential energy at the throat of the wormhole plays a significant role in making it traversable. If the gravitational potential is too strong, then the particle cannot pass the throat and becomes a black hole. Therefore, we have checked the behavior of total gravitational energy $E_g$ and shown it in Fig.~\ref{eg1}. From those profiles, one can see that the gravitational potential is comparatively high near the throat and decreases towards the surface of the wormhole. Finally, we have examined the energy conditions and presented them in Fig.~\ref{ec}. The energy density is positive throughout the configuration. Also, it is seen that NEC and SEC are violated. NEC is violated due to the highly negative radial pressure. Wormhole solutions with violation of NEC and SEC have been studied in \cite{Hochberg1998,Jusufi2020}.
These results allow us to verify the viability of different types of traversable wormhole geometries in the context of $f(R,\,T)$ gravity with the presence of conformal symmetry. Besides, it would be interesting to explore more wormhole geometries in this framework in the near future.

\section*{Acknowledgements} P.B. is thankful to the Inter University
Centre for Astronomy and Astrophysics (IUCAA), Pune, Government of India, for providing visiting associateship.

\bibliography{wormfrt}

\begin{thebibliography}{95}
\expandafter\ifx\csname natexlab\endcsname\relax\def\natexlab#1{#1}\fi
\expandafter\ifx\csname bibnamefont\endcsname\relax
  \def\bibnamefont#1{#1}\fi
\expandafter\ifx\csname bibfnamefont\endcsname\relax
  \def\bibfnamefont#1{#1}\fi
\expandafter\ifx\csname citenamefont\endcsname\relax
  \def\citenamefont#1{#1}\fi
\expandafter\ifx\csname url\endcsname\relax
  \def\url#1{\texttt{#1}}\fi
\expandafter\ifx\csname urlprefix\endcsname\relax\def\urlprefix{URL }\fi
\providecommand{\bibinfo}[2]{#2}
\providecommand{\eprint}[2][]{\url{#2}}

\bibitem[{\citenamefont{Flamm}(1916)}]{Flamm1916}
\bibinfo{author}{\bibfnamefont{L.}~\bibnamefont{Flamm}},
  \bibinfo{journal}{Physikalische Zeitschrift} \textbf{\bibinfo{volume}{17}},
  \bibinfo{pages}{448} (\bibinfo{year}{1916}).

\bibitem[{\citenamefont{Fuller and Wheeler}(1962)}]{Fuller1962}
\bibinfo{author}{\bibfnamefont{R.~W.} \bibnamefont{Fuller}} \bibnamefont{and}
  \bibinfo{author}{\bibfnamefont{J.~A.} \bibnamefont{Wheeler}},
  \bibinfo{journal}{Phys. Rev.} \textbf{\bibinfo{volume}{128}},
  \bibinfo{pages}{919} (\bibinfo{year}{1962}).

\bibitem[{\citenamefont{Einstein and Rosen}(1935)}]{Einstein1935}
\bibinfo{author}{\bibfnamefont{A.}~\bibnamefont{Einstein}} \bibnamefont{and}
  \bibinfo{author}{\bibfnamefont{N.}~\bibnamefont{Rosen}},
  \bibinfo{journal}{Phys. Rev.} \textbf{\bibinfo{volume}{48}},
  \bibinfo{pages}{73} (\bibinfo{year}{1935}).

\bibitem[{\citenamefont{Weyl}(1921)}]{Weyl1921}
\bibinfo{author}{\bibfnamefont{H.}~\bibnamefont{Weyl}},
  \bibinfo{journal}{Annalen der Physik} \textbf{\bibinfo{volume}{370}},
  \bibinfo{pages}{541} (\bibinfo{year}{1921}).

\bibitem[{\citenamefont{Wheeler}(1962)}]{Wheeler1962}
\bibinfo{author}{\bibfnamefont{J.}~\bibnamefont{Wheeler}},
  \emph{\bibinfo{title}{Geometrodynamics (New York: Academic)}},
  vol.~\bibinfo{volume}{1} (\bibinfo{publisher}{Academic Press},
  \bibinfo{year}{1962}).

\bibitem[{\citenamefont{Hawking}(1988)}]{Hawking1988}
\bibinfo{author}{\bibfnamefont{S.~W.} \bibnamefont{Hawking}},
  \bibinfo{journal}{Phys. Rev. D} \textbf{\bibinfo{volume}{37}},
  \bibinfo{pages}{904} (\bibinfo{year}{1988}).

\bibitem[{\citenamefont{Morris and Thorne}(1988)}]{Morris1988}
\bibinfo{author}{\bibfnamefont{M.~S.} \bibnamefont{Morris}} \bibnamefont{and}
  \bibinfo{author}{\bibfnamefont{K.~S.} \bibnamefont{Thorne}},
  \bibinfo{journal}{Am. J. Phys.} \textbf{\bibinfo{volume}{56}},
  \bibinfo{pages}{395} (\bibinfo{year}{1988}).

\bibitem[{\citenamefont{Visser}(1995)}]{Visser1995}
\bibinfo{author}{\bibfnamefont{M.}~\bibnamefont{Visser}},
  \emph{\bibinfo{title}{{Lorentzian wormholes: From Einstein to Hawking}}}
  (\bibinfo{year}{1995}), ISBN \bibinfo{isbn}{978-1-56396-653-8}.

\bibitem[{\citenamefont{Lemos et~al.}(2003)\citenamefont{Lemos, Lobo, and
  Quinet~de Oliveira}}]{Lemos2003}
\bibinfo{author}{\bibfnamefont{J.~P.~S.} \bibnamefont{Lemos}},
  \bibinfo{author}{\bibfnamefont{F.~S.~N.} \bibnamefont{Lobo}},
  \bibnamefont{and} \bibinfo{author}{\bibfnamefont{S.}~\bibnamefont{Quinet~de
  Oliveira}}, \bibinfo{journal}{Phys. Rev. D} \textbf{\bibinfo{volume}{68}},
  \bibinfo{pages}{064004} (\bibinfo{year}{2003}), \eprint{gr-qc/0302049}.

\bibitem[{\citenamefont{Jordan}(1959)}]{Jordan1959}
\bibinfo{author}{\bibfnamefont{P.}~\bibnamefont{Jordan}}, \bibinfo{journal}{Z.
  Phys.} \textbf{\bibinfo{volume}{157}}, \bibinfo{pages}{112}
  (\bibinfo{year}{1959}).

\bibitem[{\citenamefont{Nandi et~al.}(2006)\citenamefont{Nandi, Zhang, and
  Zakharov}}]{Nandi2006}
\bibinfo{author}{\bibfnamefont{K.~K.} \bibnamefont{Nandi}},
  \bibinfo{author}{\bibfnamefont{Y.-Z.} \bibnamefont{Zhang}}, \bibnamefont{and}
  \bibinfo{author}{\bibfnamefont{A.~V.} \bibnamefont{Zakharov}},
  \bibinfo{journal}{Phys. Rev. D} \textbf{\bibinfo{volume}{74}},
  \bibinfo{pages}{024020} (\bibinfo{year}{2006}), \eprint{gr-qc/0602062}.

\bibitem[{\citenamefont{Harko et~al.}(2009)\citenamefont{Harko, Kovacs, and
  Lobo}}]{Harko2009}
\bibinfo{author}{\bibfnamefont{T.}~\bibnamefont{Harko}},
  \bibinfo{author}{\bibfnamefont{Z.}~\bibnamefont{Kovacs}}, \bibnamefont{and}
  \bibinfo{author}{\bibfnamefont{F.~S.~N.} \bibnamefont{Lobo}},
  \bibinfo{journal}{Phys. Rev. D} \textbf{\bibinfo{volume}{79}},
  \bibinfo{pages}{064001} (\bibinfo{year}{2009}), \eprint{0901.3926}.

\bibitem[{\citenamefont{Bambi}(2013)}]{Bambi2013}
\bibinfo{author}{\bibfnamefont{C.}~\bibnamefont{Bambi}},
  \bibinfo{journal}{Phys. Rev. D} \textbf{\bibinfo{volume}{87}},
  \bibinfo{pages}{107501} (\bibinfo{year}{2013}), \eprint{1304.5691}.

\bibitem[{\citenamefont{Kuhfittig}(2014)}]{Kuhfittig2013}
\bibinfo{author}{\bibfnamefont{P.~K.~F.} \bibnamefont{Kuhfittig}},
  \bibinfo{journal}{Eur. Phys. J. C} \textbf{\bibinfo{volume}{74}},
  \bibinfo{pages}{2818} (\bibinfo{year}{2014}), \eprint{1311.2274}.

\bibitem[{\citenamefont{Rahaman et~al.}(2014)\citenamefont{Rahaman, Kuhfittig,
  Ray, and Islam}}]{Rahaman2013}
\bibinfo{author}{\bibfnamefont{F.}~\bibnamefont{Rahaman}},
  \bibinfo{author}{\bibfnamefont{P.~K.~F.} \bibnamefont{Kuhfittig}},
  \bibinfo{author}{\bibfnamefont{S.}~\bibnamefont{Ray}}, \bibnamefont{and}
  \bibinfo{author}{\bibfnamefont{N.}~\bibnamefont{Islam}},
  \bibinfo{journal}{Eur. Phys. J. C} \textbf{\bibinfo{volume}{74}},
  \bibinfo{pages}{2750} (\bibinfo{year}{2014}), \eprint{1307.1237}.

\bibitem[{\citenamefont{Li and Bambi}(2014)}]{Li2014}
\bibinfo{author}{\bibfnamefont{Z.}~\bibnamefont{Li}} \bibnamefont{and}
  \bibinfo{author}{\bibfnamefont{C.}~\bibnamefont{Bambi}},
  \bibinfo{journal}{Phys. Rev. D} \textbf{\bibinfo{volume}{90}},
  \bibinfo{pages}{024071} (\bibinfo{year}{2014}), \eprint{1405.1883}.

\bibitem[{\citenamefont{Tsukamoto}(2016)}]{Tsukamoto2016}
\bibinfo{author}{\bibfnamefont{N.}~\bibnamefont{Tsukamoto}},
  \bibinfo{journal}{Phys. Rev. D} \textbf{\bibinfo{volume}{94}},
  \bibinfo{pages}{124001} (\bibinfo{year}{2016}), \eprint{1607.07022}.

\bibitem[{\citenamefont{Zhou et~al.}(2016)\citenamefont{Zhou,
  Cardenas-Avendano, Bambi, Kleihaus, and Kunz}}]{Zhou2016}
\bibinfo{author}{\bibfnamefont{M.}~\bibnamefont{Zhou}},
  \bibinfo{author}{\bibfnamefont{A.}~\bibnamefont{Cardenas-Avendano}},
  \bibinfo{author}{\bibfnamefont{C.}~\bibnamefont{Bambi}},
  \bibinfo{author}{\bibfnamefont{B.}~\bibnamefont{Kleihaus}}, \bibnamefont{and}
  \bibinfo{author}{\bibfnamefont{J.}~\bibnamefont{Kunz}},
  \bibinfo{journal}{Phys. Rev. D} \textbf{\bibinfo{volume}{94}},
  \bibinfo{pages}{024036} (\bibinfo{year}{2016}), \eprint{1603.07448}.

\bibitem[{\citenamefont{Bhar et~al.}(2016)\citenamefont{Bhar, Rahaman, Manna,
  and Banerjee}}]{Bhar2016a}
\bibinfo{author}{\bibfnamefont{P.}~\bibnamefont{Bhar}},
  \bibinfo{author}{\bibfnamefont{F.}~\bibnamefont{Rahaman}},
  \bibinfo{author}{\bibfnamefont{T.}~\bibnamefont{Manna}}, \bibnamefont{and}
  \bibinfo{author}{\bibfnamefont{A.}~\bibnamefont{Banerjee}},
  \bibinfo{journal}{Eur. Phys. J. C} \textbf{\bibinfo{volume}{76}},
  \bibinfo{pages}{708} (\bibinfo{year}{2016}), \eprint{1612.04669}.

\bibitem[{\citenamefont{Tsukamoto}(2017)}]{Tsukamoto2017}
\bibinfo{author}{\bibfnamefont{N.}~\bibnamefont{Tsukamoto}},
  \bibinfo{journal}{Phys. Rev. D} \textbf{\bibinfo{volume}{95}},
  \bibinfo{pages}{084021} (\bibinfo{year}{2017}), \eprint{1701.09169}.

\bibitem[{\citenamefont{Morris et~al.}(1988)\citenamefont{Morris, Thorne, and
  Yurtsever}}]{Morris1988a}
\bibinfo{author}{\bibfnamefont{M.~S.} \bibnamefont{Morris}},
  \bibinfo{author}{\bibfnamefont{K.~S.} \bibnamefont{Thorne}},
  \bibnamefont{and}
  \bibinfo{author}{\bibfnamefont{U.}~\bibnamefont{Yurtsever}},
  \bibinfo{journal}{Phys. Rev. Lett.} \textbf{\bibinfo{volume}{61}},
  \bibinfo{pages}{1446} (\bibinfo{year}{1988}).

\bibitem[{\citenamefont{Frolov and Novikov}(1990)}]{Frolov1990}
\bibinfo{author}{\bibfnamefont{V.~P.} \bibnamefont{Frolov}} \bibnamefont{and}
  \bibinfo{author}{\bibfnamefont{I.~D.} \bibnamefont{Novikov}},
  \bibinfo{journal}{Phys. Rev. D} \textbf{\bibinfo{volume}{42}},
  \bibinfo{pages}{1057} (\bibinfo{year}{1990}).

\bibitem[{\citenamefont{Kar and Sahdev}(1996)}]{Kar1995}
\bibinfo{author}{\bibfnamefont{S.}~\bibnamefont{Kar}} \bibnamefont{and}
  \bibinfo{author}{\bibfnamefont{D.}~\bibnamefont{Sahdev}},
  \bibinfo{journal}{Phys. Rev. D} \textbf{\bibinfo{volume}{53}},
  \bibinfo{pages}{722} (\bibinfo{year}{1996}), \eprint{gr-qc/9506094}.

\bibitem[{\citenamefont{Arellano and Lobo}(2006)}]{Arellano2006}
\bibinfo{author}{\bibfnamefont{A.~V.~B.} \bibnamefont{Arellano}}
  \bibnamefont{and} \bibinfo{author}{\bibfnamefont{F.~S.~N.}
  \bibnamefont{Lobo}}, \bibinfo{journal}{Class. Quant. Grav.}
  \textbf{\bibinfo{volume}{23}}, \bibinfo{pages}{5811} (\bibinfo{year}{2006}),
  \eprint{gr-qc/0608003}.

\bibitem[{\citenamefont{Cataldo et~al.}(2011)\citenamefont{Cataldo, Meza, and
  Minning}}]{Cataldo2011}
\bibinfo{author}{\bibfnamefont{M.}~\bibnamefont{Cataldo}},
  \bibinfo{author}{\bibfnamefont{P.}~\bibnamefont{Meza}}, \bibnamefont{and}
  \bibinfo{author}{\bibfnamefont{P.}~\bibnamefont{Minning}},
  \bibinfo{journal}{Phys. Rev. D} \textbf{\bibinfo{volume}{83}},
  \bibinfo{pages}{044050} (\bibinfo{year}{2011}), \eprint{1101.5034}.

\bibitem[{\citenamefont{Mehdizadeh et~al.}(2015)\citenamefont{Mehdizadeh,
  Kord~Zangeneh, and Lobo}}]{Mehdizadeh2015}
\bibinfo{author}{\bibfnamefont{M.~R.} \bibnamefont{Mehdizadeh}},
  \bibinfo{author}{\bibfnamefont{M.}~\bibnamefont{Kord~Zangeneh}},
  \bibnamefont{and} \bibinfo{author}{\bibfnamefont{F.~S.~N.}
  \bibnamefont{Lobo}}, \bibinfo{journal}{Phys. Rev. D}
  \textbf{\bibinfo{volume}{92}}, \bibinfo{pages}{044022}
  (\bibinfo{year}{2015}), \eprint{1506.03427}.

\bibitem[{\citenamefont{Agnese and La~Camera}(1995)}]{Agnese1995}
\bibinfo{author}{\bibfnamefont{A.~G.} \bibnamefont{Agnese}} \bibnamefont{and}
  \bibinfo{author}{\bibfnamefont{M.}~\bibnamefont{La~Camera}},
  \bibinfo{journal}{Phys. Rev. D} \textbf{\bibinfo{volume}{51}},
  \bibinfo{pages}{2011} (\bibinfo{year}{1995}).

\bibitem[{\citenamefont{Nandi et~al.}(1997)\citenamefont{Nandi, Islam, and
  Evans}}]{Nandi1997}
\bibinfo{author}{\bibfnamefont{K.~K.} \bibnamefont{Nandi}},
  \bibinfo{author}{\bibfnamefont{A.}~\bibnamefont{Islam}}, \bibnamefont{and}
  \bibinfo{author}{\bibfnamefont{J.}~\bibnamefont{Evans}},
  \bibinfo{journal}{Phys. Rev. D} \textbf{\bibinfo{volume}{55}},
  \bibinfo{pages}{2497} (\bibinfo{year}{1997}), \eprint{0906.0436}.

\bibitem[{\citenamefont{Dzhunushaliev and Singleton}(1999)}]{Dzhunushaliev1998}
\bibinfo{author}{\bibfnamefont{V.~D.} \bibnamefont{Dzhunushaliev}}
  \bibnamefont{and}
  \bibinfo{author}{\bibfnamefont{D.}~\bibnamefont{Singleton}},
  \bibinfo{journal}{Phys. Rev. D} \textbf{\bibinfo{volume}{59}},
  \bibinfo{pages}{064018} (\bibinfo{year}{1999}), \eprint{gr-qc/9807086}.

\bibitem[{\citenamefont{Bronnikov and Grinyok}(2004)}]{Bronnikov2004}
\bibinfo{author}{\bibfnamefont{K.~A.} \bibnamefont{Bronnikov}}
  \bibnamefont{and} \bibinfo{author}{\bibfnamefont{S.~V.}
  \bibnamefont{Grinyok}}, \bibinfo{journal}{Grav. Cosmol.}
  \textbf{\bibinfo{volume}{10}}, \bibinfo{pages}{237} (\bibinfo{year}{2004}),
  \eprint{gr-qc/0411063}.

\bibitem[{\citenamefont{Richarte and Simeone}(2009)}]{Richarte2009}
\bibinfo{author}{\bibfnamefont{M.~G.} \bibnamefont{Richarte}} \bibnamefont{and}
  \bibinfo{author}{\bibfnamefont{C.}~\bibnamefont{Simeone}},
  \bibinfo{journal}{Phys. Rev. D} \textbf{\bibinfo{volume}{80}},
  \bibinfo{pages}{104033} (\bibinfo{year}{2009}), \bibinfo{note}{[Erratum:
  Phys.Rev.D 81, 109903 (2010)]}, \eprint{2006.12272}.

\bibitem[{\citenamefont{de~Leon}(2009)}]{deLeon2009}
\bibinfo{author}{\bibfnamefont{J.~P.} \bibnamefont{de~Leon}},
  \bibinfo{journal}{JCAP} \textbf{\bibinfo{volume}{11}}, \bibinfo{pages}{013}
  (\bibinfo{year}{2009}), \eprint{0910.3388}.

\bibitem[{\citenamefont{Lobo and Oliveira}(2010)}]{Lobo2010}
\bibinfo{author}{\bibfnamefont{F.~S.~N.} \bibnamefont{Lobo}} \bibnamefont{and}
  \bibinfo{author}{\bibfnamefont{M.~A.} \bibnamefont{Oliveira}},
  \bibinfo{journal}{Phys. Rev. D} \textbf{\bibinfo{volume}{81}},
  \bibinfo{pages}{067501} (\bibinfo{year}{2010}), \eprint{1001.0995}.

\bibitem[{\citenamefont{Eiroa and Figueroa~Aguirre}(2012)}]{Eiroa2012}
\bibinfo{author}{\bibfnamefont{E.~F.} \bibnamefont{Eiroa}} \bibnamefont{and}
  \bibinfo{author}{\bibfnamefont{G.}~\bibnamefont{Figueroa~Aguirre}},
  \bibinfo{journal}{Eur. Phys. J. C} \textbf{\bibinfo{volume}{72}},
  \bibinfo{pages}{2240} (\bibinfo{year}{2012}), \eprint{1205.2685}.

\bibitem[{\citenamefont{Kord~Zangeneh et~al.}(2015)\citenamefont{Kord~Zangeneh,
  Lobo, and Dehghani}}]{Zangeneh2015}
\bibinfo{author}{\bibfnamefont{M.}~\bibnamefont{Kord~Zangeneh}},
  \bibinfo{author}{\bibfnamefont{F.~S.~N.} \bibnamefont{Lobo}},
  \bibnamefont{and} \bibinfo{author}{\bibfnamefont{M.~H.}
  \bibnamefont{Dehghani}}, \bibinfo{journal}{Phys. Rev. D}
  \textbf{\bibinfo{volume}{92}}, \bibinfo{pages}{124049}
  (\bibinfo{year}{2015}), \eprint{1510.07089}.

\bibitem[{\citenamefont{Shaikh and Kar}(2016)}]{Shaikh2016}
\bibinfo{author}{\bibfnamefont{R.}~\bibnamefont{Shaikh}} \bibnamefont{and}
  \bibinfo{author}{\bibfnamefont{S.}~\bibnamefont{Kar}},
  \bibinfo{journal}{Phys. Rev. D} \textbf{\bibinfo{volume}{94}},
  \bibinfo{pages}{024011} (\bibinfo{year}{2016}), \eprint{1604.02857}.

\bibitem[{\citenamefont{Mehdizadeh and Ziaie}(2017)}]{Mehdizadeh2017}
\bibinfo{author}{\bibfnamefont{M.~R.} \bibnamefont{Mehdizadeh}}
  \bibnamefont{and} \bibinfo{author}{\bibfnamefont{A.~H.} \bibnamefont{Ziaie}},
  \bibinfo{journal}{Phys. Rev. D} \textbf{\bibinfo{volume}{95}},
  \bibinfo{pages}{064049} (\bibinfo{year}{2017}), \eprint{1704.06923}.

\bibitem[{\citenamefont{Jain et~al.}(2013)\citenamefont{Jain, Vikram, and
  Sakstein}}]{Jain2012}
\bibinfo{author}{\bibfnamefont{B.}~\bibnamefont{Jain}},
  \bibinfo{author}{\bibfnamefont{V.}~\bibnamefont{Vikram}}, \bibnamefont{and}
  \bibinfo{author}{\bibfnamefont{J.}~\bibnamefont{Sakstein}},
  \bibinfo{journal}{Astrophys. J.} \textbf{\bibinfo{volume}{779}},
  \bibinfo{pages}{39} (\bibinfo{year}{2013}), \eprint{1204.6044}.

\bibitem[{\citenamefont{Vikram et~al.}(2013)\citenamefont{Vikram, Cabr\'e,
  Jain, and VanderPlas}}]{Vikram2013}
\bibinfo{author}{\bibfnamefont{V.}~\bibnamefont{Vikram}},
  \bibinfo{author}{\bibfnamefont{A.}~\bibnamefont{Cabr\'e}},
  \bibinfo{author}{\bibfnamefont{B.}~\bibnamefont{Jain}}, \bibnamefont{and}
  \bibinfo{author}{\bibfnamefont{J.~T.} \bibnamefont{VanderPlas}},
  \bibinfo{journal}{JCAP} \textbf{\bibinfo{volume}{08}}, \bibinfo{pages}{020}
  (\bibinfo{year}{2013}), \eprint{1303.0295}.

\bibitem[{\citenamefont{Cabre et~al.}(2012)\citenamefont{Cabre, Vikram, Zhao,
  Jain, and Koyama}}]{Cabre2012}
\bibinfo{author}{\bibfnamefont{A.}~\bibnamefont{Cabre}},
  \bibinfo{author}{\bibfnamefont{V.}~\bibnamefont{Vikram}},
  \bibinfo{author}{\bibfnamefont{G.-B.} \bibnamefont{Zhao}},
  \bibinfo{author}{\bibfnamefont{B.}~\bibnamefont{Jain}}, \bibnamefont{and}
  \bibinfo{author}{\bibfnamefont{K.}~\bibnamefont{Koyama}},
  \bibinfo{journal}{JCAP} \textbf{\bibinfo{volume}{07}}, \bibinfo{pages}{034}
  (\bibinfo{year}{2012}), \eprint{1204.6046}.

\bibitem[{\citenamefont{Ade et~al.}(2016)}]{Planck2015}
\bibinfo{author}{\bibfnamefont{P.~A.~R.} \bibnamefont{Ade}}
  \bibnamefont{et~al.} (\bibinfo{collaboration}{Planck}),
  \bibinfo{journal}{Astron. Astrophys.} \textbf{\bibinfo{volume}{594}},
  \bibinfo{pages}{A14} (\bibinfo{year}{2016}), \eprint{1502.01590}.

\bibitem[{\citenamefont{Amendola et~al.}(2007)\citenamefont{Amendola, Polarski,
  and Tsujikawa}}]{Amendola2006}
\bibinfo{author}{\bibfnamefont{L.}~\bibnamefont{Amendola}},
  \bibinfo{author}{\bibfnamefont{D.}~\bibnamefont{Polarski}}, \bibnamefont{and}
  \bibinfo{author}{\bibfnamefont{S.}~\bibnamefont{Tsujikawa}},
  \bibinfo{journal}{Phys. Rev. Lett.} \textbf{\bibinfo{volume}{98}},
  \bibinfo{pages}{131302} (\bibinfo{year}{2007}), \eprint{astro-ph/0603703}.

\bibitem[{\citenamefont{Amarzguioui et~al.}(2006)\citenamefont{Amarzguioui,
  Elgaroy, Mota, and Multamaki}}]{Amarzguioui2005}
\bibinfo{author}{\bibfnamefont{M.}~\bibnamefont{Amarzguioui}},
  \bibinfo{author}{\bibfnamefont{O.}~\bibnamefont{Elgaroy}},
  \bibinfo{author}{\bibfnamefont{D.~F.} \bibnamefont{Mota}}, \bibnamefont{and}
  \bibinfo{author}{\bibfnamefont{T.}~\bibnamefont{Multamaki}},
  \bibinfo{journal}{Astron. Astrophys.} \textbf{\bibinfo{volume}{454}},
  \bibinfo{pages}{707} (\bibinfo{year}{2006}), \eprint{astro-ph/0510519}.

\bibitem[{\citenamefont{Konoplya}(2018)}]{Konoplya2018}
\bibinfo{author}{\bibfnamefont{R.~A.} \bibnamefont{Konoplya}},
  \bibinfo{journal}{Phys. Lett. B} \textbf{\bibinfo{volume}{784}},
  \bibinfo{pages}{43} (\bibinfo{year}{2018}), \eprint{1805.04718}.

\bibitem[{\citenamefont{Kuhfittig}(1999)}]{Kuhfittig1999}
\bibinfo{author}{\bibfnamefont{P.~K.~F.} \bibnamefont{Kuhfittig}},
  \bibinfo{journal}{Am. J. Phys.} \textbf{\bibinfo{volume}{67}},
  \bibinfo{pages}{125} (\bibinfo{year}{1999}).

\bibitem[{\citenamefont{Godani and Samanta}(2018)}]{Godani2018}
\bibinfo{author}{\bibfnamefont{N.}~\bibnamefont{Godani}} \bibnamefont{and}
  \bibinfo{author}{\bibfnamefont{G.~C.} \bibnamefont{Samanta}},
  \bibinfo{journal}{Int. J. Mod. Phys. D} \textbf{\bibinfo{volume}{28}},
  \bibinfo{pages}{1950039} (\bibinfo{year}{2018}), \eprint{1809.00341}.

\bibitem[{\citenamefont{Buchdahl}(1970)}]{Buchdahl1970}
\bibinfo{author}{\bibfnamefont{H.~A.} \bibnamefont{Buchdahl}},
  \bibinfo{journal}{Mon. Not. Roy. Astron. Soc.}
  \textbf{\bibinfo{volume}{150}}, \bibinfo{pages}{1} (\bibinfo{year}{1970}).

\bibitem[{\citenamefont{Starobinsky}(1980)}]{Starobinsky1980}
\bibinfo{author}{\bibfnamefont{A.~A.} \bibnamefont{Starobinsky}},
  \bibinfo{journal}{Phys. Lett. B} \textbf{\bibinfo{volume}{91}},
  \bibinfo{pages}{99} (\bibinfo{year}{1980}).

\bibitem[{\citenamefont{Thongkool et~al.}(2009)\citenamefont{Thongkool, Sami,
  Gannouji, and Jhingan}}]{Thongkool2009}
\bibinfo{author}{\bibfnamefont{I.}~\bibnamefont{Thongkool}},
  \bibinfo{author}{\bibfnamefont{M.}~\bibnamefont{Sami}},
  \bibinfo{author}{\bibfnamefont{R.}~\bibnamefont{Gannouji}}, \bibnamefont{and}
  \bibinfo{author}{\bibfnamefont{S.}~\bibnamefont{Jhingan}},
  \bibinfo{journal}{Phys. Rev. D} \textbf{\bibinfo{volume}{80}},
  \bibinfo{pages}{043523} (\bibinfo{year}{2009}), \eprint{0906.2460}.

\bibitem[{\citenamefont{Appleby et~al.}(2010)\citenamefont{Appleby, Battye, and
  Starobinsky}}]{Appleby2009}
\bibinfo{author}{\bibfnamefont{S.~A.} \bibnamefont{Appleby}},
  \bibinfo{author}{\bibfnamefont{R.~A.} \bibnamefont{Battye}},
  \bibnamefont{and} \bibinfo{author}{\bibfnamefont{A.~A.}
  \bibnamefont{Starobinsky}}, \bibinfo{journal}{JCAP}
  \textbf{\bibinfo{volume}{06}}, \bibinfo{pages}{005} (\bibinfo{year}{2010}),
  \eprint{0909.1737}.

\bibitem[{\citenamefont{Nojiri and Odintsov}(2011)}]{Nojiri2010}
\bibinfo{author}{\bibfnamefont{S.}~\bibnamefont{Nojiri}} \bibnamefont{and}
  \bibinfo{author}{\bibfnamefont{S.~D.} \bibnamefont{Odintsov}},
  \bibinfo{journal}{Phys. Rept.} \textbf{\bibinfo{volume}{505}},
  \bibinfo{pages}{59} (\bibinfo{year}{2011}), \eprint{1011.0544}.

\bibitem[{\citenamefont{Capozziello et~al.}(2018)\citenamefont{Capozziello,
  Mantica, and Molinari}}]{Capozziello2018}
\bibinfo{author}{\bibfnamefont{S.}~\bibnamefont{Capozziello}},
  \bibinfo{author}{\bibfnamefont{C.~A.} \bibnamefont{Mantica}},
  \bibnamefont{and} \bibinfo{author}{\bibfnamefont{L.~G.}
  \bibnamefont{Molinari}}, \bibinfo{journal}{Int. J. Geom. Meth. Mod. Phys.}
  \textbf{\bibinfo{volume}{16}}, \bibinfo{pages}{1950008}
  (\bibinfo{year}{2018}), \eprint{1810.03204}.

\bibitem[{\citenamefont{Sbis\`a et~al.}(2019)\citenamefont{Sbis\`a, Piattella,
  and Jor\'as}}]{Sbisa2018}
\bibinfo{author}{\bibfnamefont{F.}~\bibnamefont{Sbis\`a}},
  \bibinfo{author}{\bibfnamefont{O.~F.} \bibnamefont{Piattella}},
  \bibnamefont{and} \bibinfo{author}{\bibfnamefont{S.~E.}
  \bibnamefont{Jor\'as}}, \bibinfo{journal}{Phys. Rev. D}
  \textbf{\bibinfo{volume}{99}}, \bibinfo{pages}{104046}
  (\bibinfo{year}{2019}), \eprint{1811.01322}.

\bibitem[{\citenamefont{Elizalde et~al.}(2019)\citenamefont{Elizalde, Odintsov,
  Paul, and S\'aez-Chill\'on~G\'omez}}]{Elizalde2018}
\bibinfo{author}{\bibfnamefont{E.}~\bibnamefont{Elizalde}},
  \bibinfo{author}{\bibfnamefont{S.~D.} \bibnamefont{Odintsov}},
  \bibinfo{author}{\bibfnamefont{T.}~\bibnamefont{Paul}}, \bibnamefont{and}
  \bibinfo{author}{\bibfnamefont{D.}~\bibnamefont{S\'aez-Chill\'on~G\'omez}},
  \bibinfo{journal}{Phys. Rev. D} \textbf{\bibinfo{volume}{99}},
  \bibinfo{pages}{063506} (\bibinfo{year}{2019}), \eprint{1811.02960}.

\bibitem[{\citenamefont{Nascimento et~al.}(2019)\citenamefont{Nascimento, Olmo,
  Porfirio, Petrov, and Soares}}]{Nascimento2018}
\bibinfo{author}{\bibfnamefont{J.~R.} \bibnamefont{Nascimento}},
  \bibinfo{author}{\bibfnamefont{G.~J.} \bibnamefont{Olmo}},
  \bibinfo{author}{\bibfnamefont{P.~J.} \bibnamefont{Porfirio}},
  \bibinfo{author}{\bibfnamefont{A.~Y.} \bibnamefont{Petrov}},
  \bibnamefont{and} \bibinfo{author}{\bibfnamefont{A.~R.}
  \bibnamefont{Soares}}, \bibinfo{journal}{Phys. Rev. D}
  \textbf{\bibinfo{volume}{99}}, \bibinfo{pages}{064053}
  (\bibinfo{year}{2019}), \eprint{1812.00471}.

\bibitem[{\citenamefont{Odintsov and Oikonomou}(2019)}]{Odintsov2019}
\bibinfo{author}{\bibfnamefont{S.~D.} \bibnamefont{Odintsov}} \bibnamefont{and}
  \bibinfo{author}{\bibfnamefont{V.~K.} \bibnamefont{Oikonomou}},
  \bibinfo{journal}{Phys. Rev. D} \textbf{\bibinfo{volume}{99}},
  \bibinfo{pages}{064049} (\bibinfo{year}{2019}), \eprint{1901.05363}.

\bibitem[{\citenamefont{Chiba}(2003)}]{Chiba2003}
\bibinfo{author}{\bibfnamefont{T.}~\bibnamefont{Chiba}},
  \bibinfo{journal}{Phys. Lett. B} \textbf{\bibinfo{volume}{575}},
  \bibinfo{pages}{1} (\bibinfo{year}{2003}), \eprint{astro-ph/0307338}.

\bibitem[{\citenamefont{Nojiri and Odintsov}(2008)}]{Nojiri2007}
\bibinfo{author}{\bibfnamefont{S.}~\bibnamefont{Nojiri}} \bibnamefont{and}
  \bibinfo{author}{\bibfnamefont{S.~D.} \bibnamefont{Odintsov}},
  \bibinfo{journal}{Phys. Lett. B} \textbf{\bibinfo{volume}{659}},
  \bibinfo{pages}{821} (\bibinfo{year}{2008}), \eprint{0708.0924}.

\bibitem[{\citenamefont{Harko et~al.}(2011)\citenamefont{Harko, Lobo, Nojiri,
  and Odintsov}}]{Harko2011}
\bibinfo{author}{\bibfnamefont{T.}~\bibnamefont{Harko}},
  \bibinfo{author}{\bibfnamefont{F.~S.~N.} \bibnamefont{Lobo}},
  \bibinfo{author}{\bibfnamefont{S.}~\bibnamefont{Nojiri}}, \bibnamefont{and}
  \bibinfo{author}{\bibfnamefont{S.~D.} \bibnamefont{Odintsov}},
  \bibinfo{journal}{Phys. Rev. D} \textbf{\bibinfo{volume}{84}},
  \bibinfo{pages}{024020} (\bibinfo{year}{2011}), \eprint{1104.2669}.

\bibitem[{\citenamefont{Moraes and Correa}(2016)}]{Moraes2015a}
\bibinfo{author}{\bibfnamefont{P.~H. R.~S.} \bibnamefont{Moraes}}
  \bibnamefont{and} \bibinfo{author}{\bibfnamefont{R.~A.~C.}
  \bibnamefont{Correa}}, \bibinfo{journal}{Astrophys. Space Sci.}
  \textbf{\bibinfo{volume}{361}}, \bibinfo{pages}{91} (\bibinfo{year}{2016}),
  \eprint{1511.08160}.

\bibitem[{\citenamefont{Moraes et~al.}(2016{\natexlab{a}})\citenamefont{Moraes,
  Ribeiro, and Correa}}]{Moraes2016}
\bibinfo{author}{\bibfnamefont{P.~H. R.~S.} \bibnamefont{Moraes}},
  \bibinfo{author}{\bibfnamefont{G.}~\bibnamefont{Ribeiro}}, \bibnamefont{and}
  \bibinfo{author}{\bibfnamefont{R.~A.~C.} \bibnamefont{Correa}},
  \bibinfo{journal}{Astrophys. Space Sci.} \textbf{\bibinfo{volume}{361}},
  \bibinfo{pages}{227} (\bibinfo{year}{2016}{\natexlab{a}}),
  \eprint{1602.07159}.

\bibitem[{\citenamefont{Myrzakulov}(2012)}]{Myrzakulov2012}
\bibinfo{author}{\bibfnamefont{R.}~\bibnamefont{Myrzakulov}},
  \bibinfo{journal}{Eur. Phys. J. C} \textbf{\bibinfo{volume}{72}},
  \bibinfo{pages}{2203} (\bibinfo{year}{2012}), \eprint{1207.1039}.

\bibitem[{\citenamefont{Momeni et~al.}(2016)\citenamefont{Momeni, Moraes, and
  Myrzakulov}}]{Momeni2015}
\bibinfo{author}{\bibfnamefont{D.}~\bibnamefont{Momeni}},
  \bibinfo{author}{\bibfnamefont{P.~H. R.~S.} \bibnamefont{Moraes}},
  \bibnamefont{and}
  \bibinfo{author}{\bibfnamefont{R.}~\bibnamefont{Myrzakulov}},
  \bibinfo{journal}{Astrophys. Space Sci.} \textbf{\bibinfo{volume}{361}},
  \bibinfo{pages}{228} (\bibinfo{year}{2016}), \eprint{1512.04755}.

\bibitem[{\citenamefont{Harko}(2014)}]{Harko2014}
\bibinfo{author}{\bibfnamefont{T.}~\bibnamefont{Harko}},
  \bibinfo{journal}{Phys. Rev. D} \textbf{\bibinfo{volume}{90}},
  \bibinfo{pages}{044067} (\bibinfo{year}{2014}), \eprint{1408.3465}.

\bibitem[{\citenamefont{Moraes et~al.}(2016{\natexlab{b}})\citenamefont{Moraes,
  Arba\~nil, and Malheiro}}]{Moraes2015}
\bibinfo{author}{\bibfnamefont{P.~H. R.~S.} \bibnamefont{Moraes}},
  \bibinfo{author}{\bibfnamefont{J.~D.~V.} \bibnamefont{Arba\~nil}},
  \bibnamefont{and} \bibinfo{author}{\bibfnamefont{M.}~\bibnamefont{Malheiro}},
  \bibinfo{journal}{JCAP} \textbf{\bibinfo{volume}{06}}, \bibinfo{pages}{005}
  (\bibinfo{year}{2016}{\natexlab{b}}), \eprint{1511.06282}.

\bibitem[{\citenamefont{Zubair et~al.}(2016{\natexlab{a}})\citenamefont{Zubair,
  Abbas, and Noureen}}]{Zubair2015}
\bibinfo{author}{\bibfnamefont{M.}~\bibnamefont{Zubair}},
  \bibinfo{author}{\bibfnamefont{G.}~\bibnamefont{Abbas}}, \bibnamefont{and}
  \bibinfo{author}{\bibfnamefont{I.}~\bibnamefont{Noureen}},
  \bibinfo{journal}{Astrophys. Space Sci.} \textbf{\bibinfo{volume}{361}},
  \bibinfo{pages}{8} (\bibinfo{year}{2016}{\natexlab{a}}), \eprint{1512.05202}.

\bibitem[{\citenamefont{Shamir}(2015)}]{Shamir2015}
\bibinfo{author}{\bibfnamefont{M.~F.} \bibnamefont{Shamir}},
  \bibinfo{journal}{Eur. Phys. J. C} \textbf{\bibinfo{volume}{75}},
  \bibinfo{pages}{354} (\bibinfo{year}{2015}), \eprint{1507.08175}.

\bibitem[{\citenamefont{Zubair and Noureen}(2015)}]{Zubair2015a}
\bibinfo{author}{\bibfnamefont{M.}~\bibnamefont{Zubair}} \bibnamefont{and}
  \bibinfo{author}{\bibfnamefont{I.}~\bibnamefont{Noureen}},
  \bibinfo{journal}{Eur. Phys. J. C} \textbf{\bibinfo{volume}{75}},
  \bibinfo{pages}{265} (\bibinfo{year}{2015}), \eprint{1505.00744}.

\bibitem[{\citenamefont{Houndjo}(2012)}]{Houndjo2011}
\bibinfo{author}{\bibfnamefont{M.~J.~S.} \bibnamefont{Houndjo}},
  \bibinfo{journal}{Int. J. Mod. Phys. D} \textbf{\bibinfo{volume}{21}},
  \bibinfo{pages}{1250003} (\bibinfo{year}{2012}), \eprint{1107.3887}.

\bibitem[{\citenamefont{Mirza and Oboudiat}(2016)}]{Mirza2014}
\bibinfo{author}{\bibfnamefont{B.}~\bibnamefont{Mirza}} \bibnamefont{and}
  \bibinfo{author}{\bibfnamefont{F.}~\bibnamefont{Oboudiat}},
  \bibinfo{journal}{Int. J. Geom. Meth. Mod. Phys.}
  \textbf{\bibinfo{volume}{13}}, \bibinfo{pages}{1650108}
  (\bibinfo{year}{2016}), \eprint{1412.6640}.

\bibitem[{\citenamefont{Correa and Moraes}(2016)}]{Correa2015}
\bibinfo{author}{\bibfnamefont{R.~A.~C.} \bibnamefont{Correa}}
  \bibnamefont{and} \bibinfo{author}{\bibfnamefont{P.~H. R.~S.}
  \bibnamefont{Moraes}}, \bibinfo{journal}{Eur. Phys. J. C}
  \textbf{\bibinfo{volume}{76}}, \bibinfo{pages}{100} (\bibinfo{year}{2016}),
  \eprint{1509.00732}.

\bibitem[{\citenamefont{Zaregonbadi et~al.}(2016)\citenamefont{Zaregonbadi,
  Farhoudi, and Riazi}}]{Zaregonbadi2016}
\bibinfo{author}{\bibfnamefont{R.}~\bibnamefont{Zaregonbadi}},
  \bibinfo{author}{\bibfnamefont{M.}~\bibnamefont{Farhoudi}}, \bibnamefont{and}
  \bibinfo{author}{\bibfnamefont{N.}~\bibnamefont{Riazi}},
  \bibinfo{journal}{Phys. Rev. D} \textbf{\bibinfo{volume}{94}},
  \bibinfo{pages}{084052} (\bibinfo{year}{2016}), \eprint{1608.00469}.

\bibitem[{\citenamefont{Das et~al.}(2016)\citenamefont{Das, Rahaman, Guha, and
  Ray}}]{Das2016}
\bibinfo{author}{\bibfnamefont{A.}~\bibnamefont{Das}},
  \bibinfo{author}{\bibfnamefont{F.}~\bibnamefont{Rahaman}},
  \bibinfo{author}{\bibfnamefont{B.~K.} \bibnamefont{Guha}}, \bibnamefont{and}
  \bibinfo{author}{\bibfnamefont{S.}~\bibnamefont{Ray}}, \bibinfo{journal}{Eur.
  Phys. J. C} \textbf{\bibinfo{volume}{76}}, \bibinfo{pages}{654}
  (\bibinfo{year}{2016}), \eprint{1608.00566}.

\bibitem[{\citenamefont{Yousaf et~al.}(2016)\citenamefont{Yousaf, Bamba, and
  Bhatti}}]{Yousaf2016}
\bibinfo{author}{\bibfnamefont{Z.}~\bibnamefont{Yousaf}},
  \bibinfo{author}{\bibfnamefont{K.}~\bibnamefont{Bamba}}, \bibnamefont{and}
  \bibinfo{author}{\bibfnamefont{M.~Z. u.~H.} \bibnamefont{Bhatti}},
  \bibinfo{journal}{Phys. Rev. D} \textbf{\bibinfo{volume}{93}},
  \bibinfo{pages}{124048} (\bibinfo{year}{2016}), \eprint{1606.00147}.

\bibitem[{\citenamefont{Azizi}(2013)}]{Azizi2012}
\bibinfo{author}{\bibfnamefont{T.}~\bibnamefont{Azizi}}, \bibinfo{journal}{Int.
  J. Theor. Phys.} \textbf{\bibinfo{volume}{52}}, \bibinfo{pages}{3486}
  (\bibinfo{year}{2013}), \eprint{1205.6957}.

\bibitem[{\citenamefont{Zubair et~al.}(2016{\natexlab{b}})\citenamefont{Zubair,
  Waheed, and Ahmad}}]{Zubair2016}
\bibinfo{author}{\bibfnamefont{M.}~\bibnamefont{Zubair}},
  \bibinfo{author}{\bibfnamefont{S.}~\bibnamefont{Waheed}}, \bibnamefont{and}
  \bibinfo{author}{\bibfnamefont{Y.}~\bibnamefont{Ahmad}},
  \bibinfo{journal}{Eur. Phys. J. C} \textbf{\bibinfo{volume}{76}},
  \bibinfo{pages}{444} (\bibinfo{year}{2016}{\natexlab{b}}),
  \eprint{1607.05998}.

\bibitem[{\citenamefont{Moraes et~al.}(2017)\citenamefont{Moraes, Correa, and
  Lobato}}]{Moraes2016a}
\bibinfo{author}{\bibfnamefont{P.~H. R.~S.} \bibnamefont{Moraes}},
  \bibinfo{author}{\bibfnamefont{R.~A.~C.} \bibnamefont{Correa}},
  \bibnamefont{and} \bibinfo{author}{\bibfnamefont{R.~V.}
  \bibnamefont{Lobato}}, \bibinfo{journal}{JCAP} \textbf{\bibinfo{volume}{07}},
  \bibinfo{pages}{029} (\bibinfo{year}{2017}), \eprint{1701.01028}.

\bibitem[{\citenamefont{Zubair et~al.}(2017)\citenamefont{Zubair, Mustafa,
  Waheed, and Abbas}}]{Zubair2017}
\bibinfo{author}{\bibfnamefont{M.}~\bibnamefont{Zubair}},
  \bibinfo{author}{\bibfnamefont{G.}~\bibnamefont{Mustafa}},
  \bibinfo{author}{\bibfnamefont{S.}~\bibnamefont{Waheed}}, \bibnamefont{and}
  \bibinfo{author}{\bibfnamefont{G.}~\bibnamefont{Abbas}},
  \bibinfo{journal}{Eur. Phys. J. C} \textbf{\bibinfo{volume}{77}},
  \bibinfo{pages}{680} (\bibinfo{year}{2017}), \eprint{1709.06914}.

\bibitem[{\citenamefont{Bhatti et~al.}(2018)\citenamefont{Bhatti, Yousaf, and
  Ilyas}}]{Bhatti2018}
\bibinfo{author}{\bibfnamefont{M.~Z.} \bibnamefont{Bhatti}},
  \bibinfo{author}{\bibfnamefont{Z.}~\bibnamefont{Yousaf}}, \bibnamefont{and}
  \bibinfo{author}{\bibfnamefont{M.}~\bibnamefont{Ilyas}},
  \bibinfo{journal}{Journal of Astrophysics and Astronomy}
  \textbf{\bibinfo{volume}{39}}, \bibinfo{pages}{1} (\bibinfo{year}{2018}).

\bibitem[{\citenamefont{Sharif and Nawazish}(2019)}]{Sharif2019}
\bibinfo{author}{\bibfnamefont{M.}~\bibnamefont{Sharif}} \bibnamefont{and}
  \bibinfo{author}{\bibfnamefont{I.}~\bibnamefont{Nawazish}},
  \bibinfo{journal}{Annals Phys.} \textbf{\bibinfo{volume}{400}},
  \bibinfo{pages}{37} (\bibinfo{year}{2019}).

\bibitem[{\citenamefont{Elizalde and Khurshudyan}(2019)}]{Elizalde2019}
\bibinfo{author}{\bibfnamefont{E.}~\bibnamefont{Elizalde}} \bibnamefont{and}
  \bibinfo{author}{\bibfnamefont{M.}~\bibnamefont{Khurshudyan}},
  \bibinfo{journal}{Phys. Rev. D} \textbf{\bibinfo{volume}{99}},
  \bibinfo{pages}{024051} (\bibinfo{year}{2019}), \eprint{1812.10840}.

\bibitem[{\citenamefont{Sahoo et~al.}(2018)\citenamefont{Sahoo, Moraes, and
  Sahoo}}]{Sahoo2017}
\bibinfo{author}{\bibfnamefont{P.~K.} \bibnamefont{Sahoo}},
  \bibinfo{author}{\bibfnamefont{P.~H. R.~S.} \bibnamefont{Moraes}},
  \bibnamefont{and} \bibinfo{author}{\bibfnamefont{P.}~\bibnamefont{Sahoo}},
  \bibinfo{journal}{Eur. Phys. J. C} \textbf{\bibinfo{volume}{78}},
  \bibinfo{pages}{46} (\bibinfo{year}{2018}), \eprint{1709.07774}.

\bibitem[{\citenamefont{Heydarzade et~al.}(2015)\citenamefont{Heydarzade,
  Riazi, and Moradpour}}]{Heydarzade2014}
\bibinfo{author}{\bibfnamefont{Y.}~\bibnamefont{Heydarzade}},
  \bibinfo{author}{\bibfnamefont{N.}~\bibnamefont{Riazi}}, \bibnamefont{and}
  \bibinfo{author}{\bibfnamefont{H.}~\bibnamefont{Moradpour}},
  \bibinfo{journal}{Can. J. Phys.} \textbf{\bibinfo{volume}{93}},
  \bibinfo{pages}{1523} (\bibinfo{year}{2015}), \eprint{1411.6294}.

\bibitem[{\citenamefont{Boehmer et~al.}(2007)\citenamefont{Boehmer, Harko, and
  Lobo}}]{Boehmer2007}
\bibinfo{author}{\bibfnamefont{C.~G.} \bibnamefont{Boehmer}},
  \bibinfo{author}{\bibfnamefont{T.}~\bibnamefont{Harko}}, \bibnamefont{and}
  \bibinfo{author}{\bibfnamefont{F.~S.~N.} \bibnamefont{Lobo}},
  \bibinfo{journal}{Phys. Rev. D} \textbf{\bibinfo{volume}{76}},
  \bibinfo{pages}{084014} (\bibinfo{year}{2007}), \eprint{0708.1537}.

\bibitem[{\citenamefont{Boehmer et~al.}(2008)\citenamefont{Boehmer, Harko, and
  Lobo}}]{Boehmer2008}
\bibinfo{author}{\bibfnamefont{C.~G.} \bibnamefont{Boehmer}},
  \bibinfo{author}{\bibfnamefont{T.}~\bibnamefont{Harko}}, \bibnamefont{and}
  \bibinfo{author}{\bibfnamefont{F.~S.~N.} \bibnamefont{Lobo}},
  \bibinfo{journal}{Class. Quant. Grav.} \textbf{\bibinfo{volume}{25}},
  \bibinfo{pages}{075016} (\bibinfo{year}{2008}), \eprint{0711.2424}.

\bibitem[{\citenamefont{Kuhfittig}(2015)}]{Kuhfittig2015}
\bibinfo{author}{\bibfnamefont{P.~K.~F.} \bibnamefont{Kuhfittig}},
  \bibinfo{journal}{Annals Phys.} \textbf{\bibinfo{volume}{355}},
  \bibinfo{pages}{115} (\bibinfo{year}{2015}), \eprint{1502.02017}.

\bibitem[{\citenamefont{Kuhfittig}(2016)}]{Kuhfittig2016}
\bibinfo{author}{\bibfnamefont{P.~K.~F.} \bibnamefont{Kuhfittig}},
  \bibinfo{journal}{Int. J. Mod. Phys. D} \textbf{\bibinfo{volume}{26}},
  \bibinfo{pages}{1750025} (\bibinfo{year}{2016}), \eprint{1607.07461}.

\bibitem[{\citenamefont{Koivisto}(2006)}]{Koivisto2005}
\bibinfo{author}{\bibfnamefont{T.}~\bibnamefont{Koivisto}},
  \bibinfo{journal}{Class. Quant. Grav.} \textbf{\bibinfo{volume}{23}},
  \bibinfo{pages}{4289} (\bibinfo{year}{2006}), \eprint{gr-qc/0505128}.

\bibitem[{\citenamefont{Dadhich et~al.}(2002)\citenamefont{Dadhich, Kar,
  Mukherji, and Visser}}]{Dadhich:2001fu}
\bibinfo{author}{\bibfnamefont{N.}~\bibnamefont{Dadhich}},
  \bibinfo{author}{\bibfnamefont{S.}~\bibnamefont{Kar}},
  \bibinfo{author}{\bibfnamefont{S.}~\bibnamefont{Mukherji}}, \bibnamefont{and}
  \bibinfo{author}{\bibfnamefont{M.}~\bibnamefont{Visser}},
  \bibinfo{journal}{Phys. Rev. D} \textbf{\bibinfo{volume}{65}},
  \bibinfo{pages}{064004} (\bibinfo{year}{2002}), \eprint{gr-qc/0109069}.

\bibitem[{\citenamefont{Lynden-Bell et~al.}(2007)\citenamefont{Lynden-Bell,
  Katz, and Bicak}}]{Lynden-Bell:2007gof}
\bibinfo{author}{\bibfnamefont{D.}~\bibnamefont{Lynden-Bell}},
  \bibinfo{author}{\bibfnamefont{J.}~\bibnamefont{Katz}}, \bibnamefont{and}
  \bibinfo{author}{\bibfnamefont{J.}~\bibnamefont{Bicak}},
  \bibinfo{journal}{Phys. Rev. D} \textbf{\bibinfo{volume}{75}},
  \bibinfo{pages}{024040} (\bibinfo{year}{2007}), \bibinfo{note}{[Erratum:
  Phys.Rev.D 75, 044901 (2007)]}, \eprint{gr-qc/0701060}.

\bibitem[{\citenamefont{Katz et~al.}(2006)\citenamefont{Katz, Lynden-Bell, and
  Bicak}}]{Katz:2006uw}
\bibinfo{author}{\bibfnamefont{J.}~\bibnamefont{Katz}},
  \bibinfo{author}{\bibfnamefont{D.}~\bibnamefont{Lynden-Bell}},
  \bibnamefont{and} \bibinfo{author}{\bibfnamefont{J.}~\bibnamefont{Bicak}},
  \bibinfo{journal}{Class. Quant. Grav.} \textbf{\bibinfo{volume}{23}},
  \bibinfo{pages}{7111} (\bibinfo{year}{2006}), \eprint{gr-qc/0610052}.

\bibitem[{\citenamefont{Nandi et~al.}(2009)\citenamefont{Nandi, Zhang, Cai, and
  Panchenko}}]{Nandi:2008ij}
\bibinfo{author}{\bibfnamefont{K.~K.} \bibnamefont{Nandi}},
  \bibinfo{author}{\bibfnamefont{Y.~Z.} \bibnamefont{Zhang}},
  \bibinfo{author}{\bibfnamefont{R.~G.} \bibnamefont{Cai}}, \bibnamefont{and}
  \bibinfo{author}{\bibfnamefont{A.}~\bibnamefont{Panchenko}},
  \bibinfo{journal}{Phys. Rev. D} \textbf{\bibinfo{volume}{79}},
  \bibinfo{pages}{024011} (\bibinfo{year}{2009}), \eprint{0809.4143}.

\bibitem[{\citenamefont{Bhar}(2020)}]{Bhar:2020abv}
\bibinfo{author}{\bibfnamefont{P.}~\bibnamefont{Bhar}}, \bibinfo{journal}{Eur.
  Phys. J. Plus} \textbf{\bibinfo{volume}{135}}, \bibinfo{pages}{757}
  (\bibinfo{year}{2020}), \eprint{2105.15144}.

\bibitem[{\citenamefont{Hochberg et~al.}(1999)\citenamefont{Hochberg,
  Molina-Paris, and Visser}}]{Hochberg1998}
\bibinfo{author}{\bibfnamefont{D.}~\bibnamefont{Hochberg}},
  \bibinfo{author}{\bibfnamefont{C.}~\bibnamefont{Molina-Paris}},
  \bibnamefont{and} \bibinfo{author}{\bibfnamefont{M.}~\bibnamefont{Visser}},
  \bibinfo{journal}{Phys. Rev. D} \textbf{\bibinfo{volume}{59}},
  \bibinfo{pages}{044011} (\bibinfo{year}{1999}), \eprint{gr-qc/9810029}.

\bibitem[{\citenamefont{Jusufi et~al.}(2020)\citenamefont{Jusufi, Banerjee, and
  Ghosh}}]{Jusufi2020}
\bibinfo{author}{\bibfnamefont{K.}~\bibnamefont{Jusufi}},
  \bibinfo{author}{\bibfnamefont{A.}~\bibnamefont{Banerjee}}, \bibnamefont{and}
  \bibinfo{author}{\bibfnamefont{S.~G.} \bibnamefont{Ghosh}},
  \bibinfo{journal}{Eur. Phys. J. C} \textbf{\bibinfo{volume}{80}},
  \bibinfo{pages}{698} (\bibinfo{year}{2020}), \eprint{2004.10750}.

\end{thebibliography}

\end{document}